\documentclass[a4paper,fleqn]{cas-sc}

\usepackage[numbers]{natbib}

\begin{document}

	\let\WriteBookmarks\relax
	\def\floatpagepagefraction{1}
	\def\textpagefraction{.001}
	\shorttitle{Cauchy-Paul wavelet revisited}
	\shortauthors{P. Argoul et~al.}

	\title[mode = title]{Cauchy-Paul wavelet transforms revisited: A framework for intermittent non-sinusoidal oscillations} 
	
	
	\author{Pierre Argoul} 
	
	\affiliation{organization={LVMT, Ecole Nationale des Ponts et Chaussées, Université Gustave Eiffel},
		city={Marne la Vallée},
		postcode={77454}, 
		country={France}}
	
	\author{Jacques Taillard} 
	
	\affiliation{organization={CNRS USR3413 SANPSY Sommeil, Addiction et NeuroPSYchiatrie, Université de Bordeaux},
		city={Bordeaux},
		postcode={33000}, 
		country={France}}
	
	\author{Fran\c{c}oise Argoul} 
	
	\affiliation{organization={LOMA, CNRS UMR5798, Universit\'e de Bordeaux},
		city={Talence},
		postcode={33405}, 
		country={France}}

	\begin{abstract}
		This paper revisits the continuous wavelet transform framework by establishing a rigorous physical and dimensional formulation of the Cauchy-Paul mother wavelet, tailored specifically for intermittent, non-sinusoidal electrophysiological oscillations. Departing from conventional, purely mathematical definitions, we introduce a characteristic time scale $\tau$ into the frequency-domain formulation of the mother wavelet. This parameter ensures strict dimensional consistency by maintaining dimensionless functional arguments, thereby confining the physical dimension solely to the multiplicative normalization constant under both $L^1(\mathbb{R})$ and $L^2(\mathbb{R})$ norms. A sharp dimensional and structural analysis of the resulting wavelet is conducted. We demonstrate that the spectral asymmetry inherent to the Cauchy-Paul wavelet dictates a strict mathematical hierarchy between three alternative reference frequencies: the peak ($L^\infty$), the centroid ($L^1$), and the energy-weighted ($L^2$) frequencies. Each frequency definition yields distinct quality factors ($Q$) and time-bandwidth characteristics that govern the time-frequency localization trade-off. To track multi-component EEG sleep pattern signals, a phase-based algebraic estimator is deployed alongside an advanced ridge-extraction method. The robust tracking performance and morphological adaptability of the proposed Cauchy-Paul framework are first numerically validated on synthetic transients, harmonics, and chirps, and subsequently applied to real, non-stationary EEG recordings to successfully isolate and decipher the non-sinusoidal signatures of sleep spindles.
	\end{abstract}
	\begin{highlights}
		\item  Advanced mathematical foundations of the Cauchy-Paul wavelet transform. 
		\item  A CWT phase-based algebraic estimator for complex non-stationary signals. 
		\item Tracking the instantaneous frequency of transient non-sinusoidal signals.
		\item Accurate analysis of intermittent and non-sinusoidal sleep spindles.
	\end{highlights}
	
	\maketitle

	\section{Introduction}
	Any waveform deviating from a pure sinusoid is termed non-sinusoidal (e.g., square, sawtooth, or pulses). These signals are ubiquitous in complex dynamical systems—from seismic bursts to cardiac cycles—where their specific morphology (sharpness, slope) encodes vital information about the underlying physical mechanisms. In electrophysiology, neural oscillations are rarely perfectly sinusoidal; instead, they exhibit asymmetric peaks and sharp transients that serve as critical markers of cortical excitability and the excitation-inhibition balance.
	
	Intermittent signals are characterized by discrete bursts of high-energy activity separated by periods of relative quiescence or low-level noise. Unlike stationary processes, which maintain constant statistical properties over time, intermittency implies a non-stationary and transient nature. This temporal fragmentation makes traditional global analysis methods, such as the Fourier Transform, inadequate for their characterization as they average out the specific dynamics of the active phases.

	The analysis of physiological signals, particularly electroencephalograms (EEG), remains a cornerstone of modern neuroscience. A primary example of such activity are the sleep spindles; short, intermittent bursts of 11--16~Hz oscillations occurring during non-rapid eye movement (NREM) sleep \cite{fernandez_sleep_2020,sitnikova_sleep_2009,de_gennaro_sleep_2003}. Beyond their role as critical biomarkers for memory consolidation, cortical plasticity, age-related, and neurodegenerative decline, spindles are fundamentally characterized by their non-sinusoidal morphology. Consequently, solving their precise onset, duration and intra-spindle waveform \cite{cole_brain_2017} provides a high diagnostic value to assess the underlying dynamics of thalamocortical networks.
	
	Recent electrophysiological studies demonstrate that neural rhythms naturally deviate from pure sinusoids, frequently exhibiting asymmetric morphologies and sharp transient profiles. These non-sinusoidal properties are now recognized as pivotal physiological markers, directly mapping specific states of cortical excitability and the temporal balance between excitation and inhibition.
	The non-sinusoidal nature of brain oscillations would require a mathematical framework well-suited for extracting waveform asymmetry.
	
	Over the past fifteen years, the use of time-frequency techniques has steadily expanded to improve both the detection and boundary analysis of sleep spindles. 
	In its early stages, Andrillon et al. \cite{andrillon_sleep_2011} proposed in 2011 a conventional sigma-band pass-band filtering coupled with envelope extraction (via Hilbert transform or amplitude smoothing) and automated thresholding to isolate spindle boundaries. 
	
	Initially, spindle detection relied on the Short-Time Fourier Transform (STFT) to extract time-frequency features; however, as reviewed by Prerau et al. \cite{prerau_sleep_2017}, this traditional approach suffers from rigid resolution limits inherent to fixed Fourier windows. This has prompted the development of non-linear sparse time-frequency representations \cite{parekh_sleep_2014} and advanced wavelet-based sparse optimization techniques \cite{parekh_detection_2015} designed to isolate the complex dynamics of these oscillations. Alternatively, the Continuous Wavelet Transform (CWT) using the Morlet wavelet was successfully introduced by Zygierewicz et al. \cite{zygierewicz_high_1999} to resolve the fine, transient structure of sleep spindles. Their work demonstrated the clear superiority of multi-resolution analysis over fixed-window Fourier transforms for capturing subtle intra-spindle variations through an adjustable time-frequency resolution.

	Indeed, the comprehensive benchmark by Warby et al. \cite{warby_sleep-spindle_2014} highlighted low human expert consensus and underscored the limitations of traditional automated detection.
	By tracking instantaneous spindle frequencies with a Morlet CWT in both epileptic and healthy rats, Sitnikova et al. \cite{sitnikova_time-frequency_2014} (2014) revealed an ascending intra-spindle frequency profile unique to the non-epileptic control group. Tsanas and Clifford (2015) \cite{tsanas_stage-independent_2015} provided a mathematical upgrade of previous Morlet based CWT by thresholding the CWT coefficients at the spindle
	frequency range, using a moving average of 100 ms sliding window.
	The multitaper approach (proposed by Dimitrov and Prerau, 2021) \cite{dimitrov_sleep_2021} averaged orthogonal windows to reduce noise, thereby revealing transient oscillatory events hidden in raw EEG tracks.

	The inherent ability of the CWT to capture transient intermittency is fundamental; unlike Fourier-based approaches that average spectral content over long temporal windows, the CWT can detect time-varying quantities with its translation parameter $b$ and isolate the waxing and waning phases of individual spindles.

	However, these conventional methods often impose a Gaussian-defined, sinusoidal symmetry that often struggles to isolate non-linear features \cite{cole_brain_2017}. Because EEG spindles exhibit nonsinusoidal morphologies, temporal asymmetries, and low-frequency biases reflecting the non-linear dynamics of thalamocortical loops, symmetric approaches inevitably mask essential physiological information.
	
	Moving beyond fixed-size multitaper STFT windowing \cite{dimitrov_sleep_2021,stokes_transient_2023}, which was recently used to highlight the inter-individual variability of spindle time-frequency patterns, the Cauchy-Paul CWT, proposed in this study, may provide moreover a specific alternative to isolates these transient events while preserving their intrinsic asymmetric 'rise-and-decay' morphology. While Morlet represented a major step forward from fixed window size filtering, the Cauchy-Paul framework  would likely push further  this advantage by matching the asymmetric, non-sinusoidal signature of sleep spindles.
	
	Furthermore, adjusting the wavelet's quality factor $Q$—as formalized by Le and Argoul \cite{le_continuous_2004} via the structural order $m$ for Cauchy-Paul wavelets allows the wavelet envelope to adapt directly to the spindle's specific waveform. This ensures also that the estimated instantaneous frequency reflects the true underlying neural pacing rather than a smeared approximation.
	
	In this work, we revisit the Cauchy-Paul wavelet framework as a robust alternative for characterizing these non-stationary signals. Because it is strictly analytic, i.e. possessing zero energy at negative frequencies, this framework allows for an unambiguous estimation of instantaneous phase and amplitude, eliminating the phase distortions inherent to non-analytic operators \cite{flandrin1999time, bruns2004fourier, cole2019cycle,munoz_continuous_2005}. Unlike varied mathematical definitions found in the literature, we endow our mother wavelet with a physical dimension through strict normalization, an internal time parameter $\tau$, and an adjustable structural order $m$. 
	
	This specific implementation offers two key advantages for tracking complex EEG transients. First, its analytical precision ensures rigorous extraction of spindle phase and envelope, identifying fine-grained features like intra-spindle ``chirps''. Second, the order $m$ acts as a tuning parameter for the time-frequency trade-off, optimizing the wavelet quality factor for either sharp temporal localization or enhanced spectral selectivity. 
	
	By decoupling the shape analysis (controlled by $m$) from the frequency tuning (controlled by $\tau$), this framework provides a robust mathematical "scalpel" to dissect the transient dynamics of the sleeping brain. In this paper, we show that the sensitivity of the Cauchy-Paul wavelet to phase curvature is well suited for detecting subtle changes in spindle morphology long before they become apparent through traditional power-spectral density analysis.

	The remainder of this paper is organized as follows. Section~2 revisits the Cauchy-Paul mother wavelet, introducing a dimensionally consistent framework from a structural order $m$ and a characteristic time scale parameter $\tau$, to set the wavelet's center frequency. Section~3 details the selection of the normalization factor in the continuous wavelet transform (CWT) definition, to toggle between $L^1(\mathbb{R})$ and $L^2(\mathbb{R})$ norms, enabling either direct amplitude or energy conservation, and recalls classical instantaneous frequency estimation techniques for mono-component signals. Section~4 extends this to multi-component signals using Cauchy-Paul continuous wavelet transform ridge-based extraction and algebraic estimators. Section~5 validates the performance of the phase estimator on various synthetic mono- and multi-component signals, including linear and non-linear chirps. Section~6 demonstrates its neurophysiological application by isolating transient sleep spindles from non-stationary EEG backgrounds. Finally, Section~7 revisits the local spectral resolution, examines the emergence of invariant phase velocity plateaus across different wavelet orders, and concludes the paper.

	\section{The Cauchy-Paul mother wavelet}
	
	The simplest complex wavelets are the Cauchy-Paul and Morlet wavelets, which are extensively used in quantum mechanics, analytic models, and signal analysis \cite{grossmann_decomposition_1984, mallat_wavelet_2009, cohen_better_2019, argoul_instantaneous_2003, le_continuous_2004}. Belonging to the broader family of Morse analytical wavelets, the Cauchy-Paul functions are also frequently referred to as Klauder wavelets (named after John R. Klauder) \cite{klauder1999wavelets} in the context of coherent states and continuous wavelet transforms.
	
	These wavelets are defined in the frequency domain with an explicit physical dimension by introducing a characteristic time scale $\tau > 0$ (where $[\tau]=T$, and $[X]$ denotes the dimensional formula of $X$). This parameter ensures that the arguments of both the power and exponential functions remain dimensionless, maintaining physical consistency, while the order parameter $m$ dictates the number of oscillations and vanishing moments. Their definition is expressed as follows:
	\begin{equation}
		\hat{\psi}_{m,\tau}(f) = A_{m,\tau} \, (f \tau)^m e^{-f \tau} H(f)\label{def-ondelette Cauchy-Paul domaine frequentiel} \;,
	\end{equation} 
	where $H(f)$ is the Heaviside step function ensuring analyticity, and $A_{m,\tau}$ is a normalization constant. Since all other factors in \eqref{def-ondelette Cauchy-Paul domaine frequentiel} are dimensionless, the dimension of the wavelet transform $\hat{\psi}_{m,\tau}(f)$ is entirely carried by its multiplicative constant: $\left[\hat{\psi}_{m,\tau}\right] = \left[ A_{m,\tau} \right]$, which adapts to the chosen normalization.
	
	\subsection{Admissibility condition}
	
	The admissibility condition ensures that the CWT is invertible, requiring the convergence of:
	\begin{equation*}
		c_{\psi} = \int_{0}^{\infty} \frac{|\hat{\psi}(f)|^2}{f} df < \infty \;.
	\end{equation*}
	
	This imposes two properties. First, the ``zero mean property'' ($\int_{-\infty}^{\infty} \psi(t) dt = 0$), implying $\hat{\psi}(0)=0$, which confirms that the wavelet acts as a band-pass filter. Second, the ``normalization'' condition for energy conservation and signal reconstruction,  $c_{\psi_{m,\tau}}$ is defined for the Cauchy-Paul  wavelet as:
	\begin{equation*}
		c_{\psi_{m,\tau}} = \frac{A_{m,\tau}^2}{2^{2m}} \Gamma(2m) \;,
	\end{equation*}
	where $\Gamma$ is the Euler Gamma function, which evaluates to $\Gamma(2m+1) = (2m)!$ for integer values of $m$.
	
	The complete mathematical derivation of this result, which confirms its dependence on both the normalization convention and the wavelet parameters, is provided in the supplementary material  \textbf{S1}. 
	
	\subsection{Time-domain derivation of the Cauchy-Paul mother wavelet}
	
	The time-domain derivation of the Cauchy-Paul mother wavelet is detailed in the supplementary material \textbf{S2} and gives the expression:
	\begin{equation}
		\psi_{m,\tau}(t) = A_{m,\tau} \tau^{m} \frac{m!}{(\tau - 2i\pi t)^{m+1}} \;.
	\end{equation}
	
	The analytic form of this wavelet, characterized by its holomorphy, enables simultaneous extraction of signal amplitude and instantaneous phase. Furthermore, the parameter $m$ controls the temporal decay, balancing the trade-off between temporal localization and frequency resolution.
	The values and physical dimensions of $A_{m,\tau}$ , $c_{\psi_{m,\tau}}$ and 
	$\psi_{m,\tau}(0)$ are summarized in Table \ref{tab:cauchy_parameters_dim} according to the chosen normalizations $L^1(\mathbb{R})$ or $L^2(\mathbb{R})$. Details of the computations are given in the supplementary material \textbf{S3}.

	\begin{table*}[h]
		\centering
		\renewcommand{\arraystretch}{2.5} 
		\begin{tabular}{|l|c|c||c|c|}
			\hline
			\textbf{Property / Parameter} & \textbf{$L^1(\mathbb{R})$ Normalization} & \textbf{Dim. $L^1$} & \textbf{$L^2(\mathbb{R})$ Normalization} & \textbf{Dim. $L^2$} \\ \hline
			\textbf{Condition} & $\displaystyle \int_{0}^{\infty} |\hat{\psi}_{m,\tau}| df = 1$ & - & $\displaystyle \int_{0}^{\infty} |\hat{\psi}_{m,\tau}|^2 df = 1$ & - \\ \hline
			\textbf{Amplitude $A_{m,\tau}$} & $\displaystyle \frac{\tau}{m!}$ & $T$ & $\displaystyle \sqrt{\frac{2^{2m+1}\tau}{(2m)!}}$ & $T^{1/2}$ \\ \hline
			\textbf{Admissibility $c_{\psi_{m,\tau}}$} & $\displaystyle \frac{\tau^2 (2m-1)!}{2^{2m} (m!)^2}$ & $T^2$ & $\displaystyle \frac{\tau}{m}$ & $T$ \\ \hline
			\textbf{Value $\psi_{m,\tau}(0)$} & $1$ & $1$ & $\displaystyle \sqrt{\frac{2^{2m+1}(m!)^2}{\tau (2m)!}}$ & $T^{-1/2}$ \\ \hline
		\end{tabular}
		\caption{Comparison of Cauchy-Paul wavelet parameters and physical dimensions according to the $L^1$ and $L^2$ norms for the mother wavelet ($T$ denotes the dimension of Time).}
		\label{tab:cauchy_parameters_dim}
	\end{table*}
	
	\subsection{Peak and mean frequencies of the mother wavelet}
	Three characteristic frequencies associated with the wavelet $\psi_{m,\tau}$, namely the ``peak frequency'' $f_{\psi_{m,\tau}}^{[0]}$, the $L^{1}(\mathbb{R})$ ``centroid frequency'' $f_{\psi_{m,\tau}}^{[1]}$ and the ``energy frequency'' $f_{\psi_{m,\tau}}^{[2]}$ are explicitly calculated in the supplementary material \textbf{S4}
	
	and presented in Table \ref{tab:cauchy_final}.
	The centroid frequency, also called the center frequency, is defined by the first moment of the spectral energy density:
	\begin{equation*}
		f_{\psi_{m,\tau}}^{[1]} = \frac{\int_{0}^{\infty} f |\hat{\psi}_{m,\tau}(f)|^2 df}{\int_{0}^{\infty} |\hat{\psi}_{m,\tau}(f)|^2 df}
		\;.
	\end{equation*}
	These three frequencies follow a specific order determined by the positive  skewness of the Cauchy-Paul wavelet $\hat{\psi}_{m,\tau}(f) $ (i.e., a sharp low-frequency rise followed by a long high-frequency tail):
	$f_{\psi_{m,\tau}}^{[0]} < f_{\psi_{m,\tau}}^{[1]} < f_{\psi_{m,\tau}}^{[2]} \;.$
	
	The peak frequency $f_{\psi_{m,\tau}}^{[0]}$ (the mode) is shifted left of the mean because the tail extends to the right. The centroid frequency $f_{\psi_{m,\tau}}^{[1]}$ (the center of gravity) is pulled rightward by this positive skew. Finally, the energy-weighted frequency $f_{\psi_{m,\tau}}^{[2]}$, derived from the second moment of $|\hat{\psi}_{m,\tau}|^2$, assigns a greater weight to higher frequencies, placing it farther into the tail.

	The spectral bandwidth is often defined as the standard deviation of the spectral energy density. This is directly linked to the Heisenberg Uncertainty Principle. If we treat the normalized energy density as a probability distribution $P(f) = \frac{|\hat{\psi}_{m,\tau}(f)|^2}{\int |\hat{\psi}_{m,\tau}(f)|^2 df}$, the corresponding RMS bandwidth or the spectral standard deviation $\Delta f_{\psi}^{[c]}$ is twice the square root of the variance $\sigma^2_{f_{\psi}^{[c]}}$ :
	\begin{equation}
		\Delta f^{[c]}_{\psi_{m,\tau}} = 2 \sqrt{\sigma^2_{f_{\psi_{m,\tau}}^{[c]}}} = 2 \sqrt{ \int_{0}^{+\infty} \left(f - f_{\psi_{m,\tau}}^{[c]}\right)^2 P(f) \, df }.
		\label{def standard deviations}
	\end{equation}
	In the following, the exponent $[c] \in \{0, 1, 2\}$ follows the conventions introduced previously, mapping the characteristic frequencies ($f_{\psi_{m,\tau}}^{[0]}$, $f_{\psi_{m,\tau}}^{[1]}$, and $f_{\psi_{m,\tau}}^{[2]}$) to their corresponding spectral bandwidths ($\Delta f_{\psi_{m,\tau}}^{[0]}$, $\Delta f_{\psi_{m,\tau}}^{[1]}$, and $\Delta f_{\psi_{m,\tau}}^{[2]}$) (see Table \ref{tab:cauchy_final}). These metrics quantify the spectral spread relative to each reference frequency, reflecting how the choice of center impacts the observed bandwidth of the Cauchy-Paul wavelet. Full computational derivations for these standard deviations, based on the second-order moments of the spectral density, are provided in the supplementary material \textbf{S5}.

	\subsection{The effective time duration of the mother wavelet}
	The effective time duration (or bandwidth) $\Delta t_{\psi_{m,\tau}}$ of the Cauchy-Paul mother wavelet, which characterizes its temporal support, is defined as twice the temporal standard deviation $\sigma_t$ to ensure consistency with the spectral bandwidth definition ($\Delta f = 2\sigma_f$):
	\begin{equation}
		\Delta t_{\psi_{m,\tau}} = 2 \sigma_t = 2 \sqrt{ \int_{-\infty}^{+\infty} (t - t_{\psi_{m,\tau}})^2 P(t) \, dt} \;,
		\label{eq:def_delta_t}
	\end{equation}
	where $P(t) = |\psi_{m,\tau}(t)|^2 / \int_{-\infty}^{+\infty} |\psi_{m,\tau}(t)|^2 \, dt$ is the normalized energy density function. 
	
	This duration is finally expressed as:
	\begin{equation}
		\Delta t_{\psi_{m,\tau}} = \tau \sqrt{\frac{2}{2m-1}} \;,
		\label{eq:final_delta_t_value}
	\end{equation}
	with $m > \frac{1}{2}$ (see  the supplementary material \textbf{S6} 
	for details on the derivation).
	
	The parameter $\Delta t_{\psi_{m,\tau}}$ characterizes the temporal resolution of the wavelet. As established by \eqref{eq:final_delta_t_value}, a high order $m$ yields a small $\Delta t_{\psi_{m,\tau}}$, providing the sharp temporal precision necessary for detecting transients or abrupt changes. Conversely, a low order $m$ results in a large $\Delta t_{\psi_{m,\tau}}$; while this reduces temporal localized precision, it is better suited for characterizing stable, long-term oscillations due to the resulting gain in frequency resolution.
	
	Metaphorically, a high order $m$ acts like a scalpel: it pinpoints the exact timing of an event with high precision but remains relatively "blind" to subtle frequency nuances. In contrast, a low order $m$ functions like a blurred magnifying glass: it enables high-precision identification of the signal's frequency content, yet obscures the precise temporal boundaries of the event on the timeline.

	
	\subsection{Time-frequency resolution and quality factor}
	
	To quantify the trade-off between temporal and spectral resolution, literature often introduces the effective number of temporal oscillations, $N^{[c]}_{\text{osc}}$ \cite{lilly_element_2017, cohen_better_2019}. This dimensionless parameter describes the wave packet's structure within its compact support and is defined as the product of the wavelet duration $\Delta t_{\psi_{m,\tau}}$ and its central frequency $f^{[c]}_{\psi_{m,\tau}}$:
	\begin{equation*}
		N^{[c]}_{\text{osc}} = \Delta t_{\psi_{m,\tau}} \, f^{[c]}_{\psi_{m,\tau}} \;,
	\end{equation*}
	where the index $[c] \in \{0, 1, 2\}$ refers to the specific central frequency definition chosen. Adjusting $N_{\text{osc}}$ optimizes the energy concentration along the spectral ridges and minimizes noise-induced fluctuations, thereby facilitating the separation of concurrent chirps.
	
	In physics and signal processing, however, the dimensionless quality factor $Q^{[c]}$ is generally preferred to characterize the frequency selectivity of a resonator or an analysis filter. It is defined as the ratio of the central frequency to the spectral bandwidth $\Delta f_{\psi_{m,\tau}}^{[c]}$ introduced in \eqref{def standard deviations}:
	\begin{equation}
		Q^{[c]} = \frac{f_{\psi_{m,\tau}}^{[c]}}{\Delta f_{\psi_{m,\tau}}^{[c]}} \;.
		\label{eq:Qc}
	\end{equation}

	A high $Q^{[c]}$ reflects narrow spectral localization and sustained temporal oscillations, whereas a low $Q^{[c]}$ denotes a broad response with rapid temporal decay.
	
	Consequently, three distinct quality factors can be deduced:\begin{equation}Q^{[0]} = \frac{f_{\psi_{m,\tau}}^{[0]}}{\Delta f_{\psi_{m,\tau}}^{[0]}},\; Q^{[1]} = \frac{f_{\psi_{m,\tau}}^{[1]}}{\Delta f_{\psi_{m,\tau}}^{[1]}}, \; Q^{[2]} = \frac{f_{\psi_{m,\tau}}^{[2]}}{\Delta f_{\psi_{m,\tau}}^{[2]}} \;.
		\label{eq:three_quality_factors}
	\end{equation}

	In wavelet theory, $Q^{[c]}$ specifically quantifies the time-frequency resolution trade-off. A fundamental property of the wavelet transform is that it maintains a constant $Q^{[c]}$ across all scales, ensuring scale-invariant localization.

	The parameters \textbf{$f^{[c]}_{\psi_{m,\tau}}$},    \textbf{$(\sigma_f^{[c]})^2$}, \textbf{$\Delta f^{[c]}_{\psi_{m,\tau}} (= 2\sigma_f^{[c]})$}, \textbf{$Q^{[c]}$} and \textbf{$\Delta t_{\psi_{m,\tau}}$} are collected in Table \ref{tab:cauchy_final}.

	\begin{table*}[htbp]
		\centering
		\caption{Parameters of the Cauchy-Paul wavelet with $\Delta_m = (m+1) - \sqrt{(m+1)(m+0.5)}$. All spectral parameters including $N_{osc}^{[c]}$ depend on the frequency reference, while the time duration remains constant.}
		\label{tab:cauchy_final}
		\vskip 0.15in
		\begin{small}
			\begin{tabular}{lcccccc}
				\toprule
				\textbf{Ref.} & \textbf{Central Freq.} & \textbf{Spect. Var.} & \textbf{Spect. Band.} & \textbf{Q-Factor} & \textbf{Time Dur.} & \textbf{Oscillations} \\ 
				($c$) & $f^{[c]}_{\psi_{m,\tau}}$ & $(\sigma_f^{[c]})^2$ & $\Delta f^{[c]}_{\psi_{m,\tau}}$ & $Q^{[c]}$ & $\Delta t_{\psi_{m,\tau}}$ & $N_{osc}^{[c]} = \Delta t_{\psi_{m,\tau}} f_{\psi_{m,\tau}}^{[c]}$ \\
				\midrule
				\textbf{Peak} ($c=0$) & 
				\vspace{0.2cm} $\frac{m}{\tau}$ & $\frac{m+1}{2\tau^2}$ & $\frac{\sqrt{2(m+1)}}{\tau}$ & $\frac{m}{\sqrt{2(m+1)}}$ & $\tau \sqrt{\frac{2}{2m-1}}$ & $m \sqrt{\frac{2}{2m-1}}$ \\ 
				
				\textbf{Centroid} ($c=1$) & 
				\vspace{0.2cm} $\frac{m+0.5}{\tau}$ & $\frac{2m+1}{4\tau^2}$ & $\frac{\sqrt{2m+1}}{\tau}$ & $\frac{1}{2}\sqrt{2m+1}$ & $\tau \sqrt{\frac{2}{2m-1}}$ & $(m+0.5) \sqrt{\frac{2}{2m-1}}$ \\ 
				
				\textbf{Energy} ($c=2$) & 
				$\frac{\sqrt{(m+1)(m+0.5)}}{\tau}$ & $\frac{(2m+1)\Delta_m}{2\tau^2}$ & $\frac{\sqrt{2(2m+1)\Delta_m}}{\tau}$ & $\sqrt{\frac{(m+1)(m+0.5)}{2(2m+1) \Delta_m}}$ & $\tau \sqrt{\frac{2}{2m-1}}$ & $\sqrt{\frac{2(m+1)(m+0.5)}{2m-1}}$ \\ 
				\bottomrule
			\end{tabular}
		\end{small}
	\end{table*}
	The quality factor $Q^{[c]}$ of the Cauchy-Paul wavelet depends on the chosen centering method. Given the convention $\Delta f^{[c]} = 2\sigma_f^{[c]}$, the relationship between the quality factors for the peak ($Q^{[0]}$), centroid ($Q^{[1]}$) and energy ($Q^{[2]}$) methods for $m \geq 1$, can be established as:
	\begin{equation}
		Q^{[0]} < Q^{[1]} < Q^{[2]}\;.
		\label{inequality Q^{[c]}}
	\end{equation}
	
	This inequality \eqref{inequality Q^{[c]}} stems from the asymmetric profile of the Cauchy-Paul spectrum. Compared to the peak ($c=0$), the centroid method ($c=1$) shifts the central frequency upward due to the heavy high-frequency tail. Meanwhile, the energy method ($c=2$) yields the narrowest effective bandwidth relative to its center frequency, resulting in a sharper spectral selectivity and a higher quality factor. 
	This ordering demonstrates that the centroid ($L^1$) method tracks the true center of gravity of the spectral asymmetric distribution, whereas the energy reference ($L^2$) yields a higher quality factor but remains highly sensitive to the high-frequency tail of the Cauchy-Paul distribution. As a consequence, the choice of the convention directly impacts the interpretation of the wavelet's selectivity.

	\begin{figure}[]
		\centering
		\includegraphics[width=0.6\textwidth]{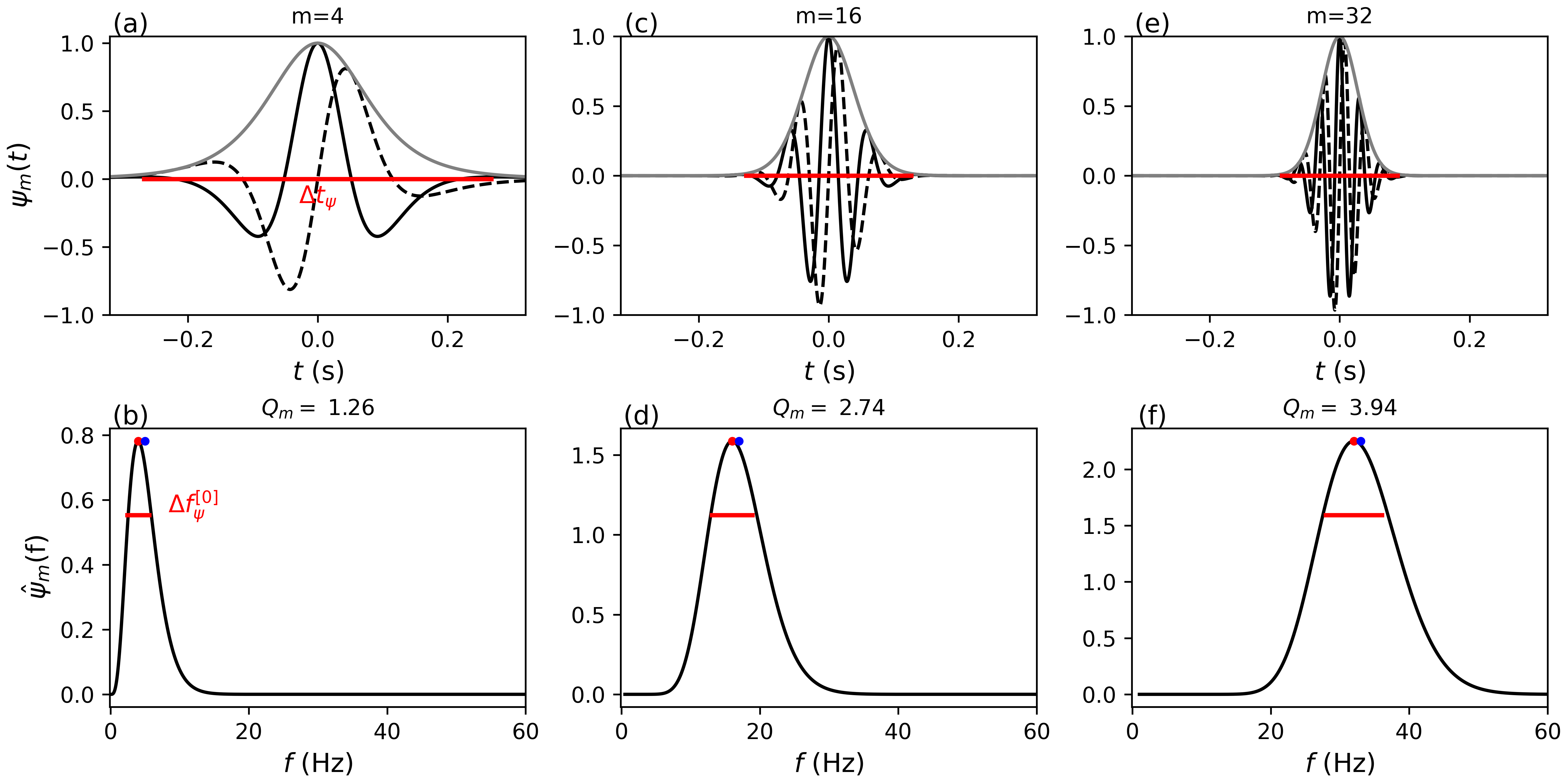}
		\vspace{-0.5em}
		\caption{Representations of the Cauchy-Paul mother wavelet in time $\psi_{m}(t)$ and in frequency $ \hat{\psi}_{m}(f)$ for increasing values of $m$ and $\tau=1$. 
			Gray lines:  $\left|\psi_{m}(t)\right|$, black lines: $\Re(\psi_{m}(t))$, dashed lines: $\Im(\psi_{m}(t))$. The temporal $\Delta t_{\psi_{m}}$ and spectral $\Delta f_{\psi_{m}}^{[0]}$ bandwidths are reported with red lines. The red (blue) dots correspond to $f_{\psi_{m}}^{[0]}$ and $f_{\psi_{m}}^{[1]}$ respectively. The $Q$ values are computed with the relation $Q^{[0]} = f_{\psi_{m}}^{[0]}/\Delta f_{\psi_{m}}^{[0]}$. }
		\label{fig:Paul_wavelet_scaled_t_f} 
	\end{figure}

	Fig. \ref{fig:Paul_wavelet_scaled_t_f}  illustrates the temporal representations $\psi_{m}(t)$ and the frequency-domain representations $\hat{\psi}_{m}(f)$ of the mother wavelet to increase $m$ values for $m\in\left\{4,16,32\right\}$. The spectrum $\hat{\psi}_{m}(f)$ peaks at $f=\frac{m}{\tau}$. It rises polynomially to the left of the peak (steep slope) and decreases exponentially to the right (gentler slope). This figure shows how the adjustment of the parameter $m$ modifies the morphology of the wavelet, providing the necessary "morphological adaptability" to accurately capture EEG signal transients.
	In Fig. \ref{fig:Paul_wavelet_scaled_t_f}, the red lines are used to illustrate the characteristic bandwidths of the Cauchy-Paul mother wavelet, namely the temporal bandwidth $\Delta t_{\psi}$ in the time-domain  and the spectral bandwidth $\Delta f_{\psi}^{[0]}$ in the frequency-domain for each order $m$.
	
	\begin{figure}[]
		\centering
		\includegraphics[width=0.6\textwidth]{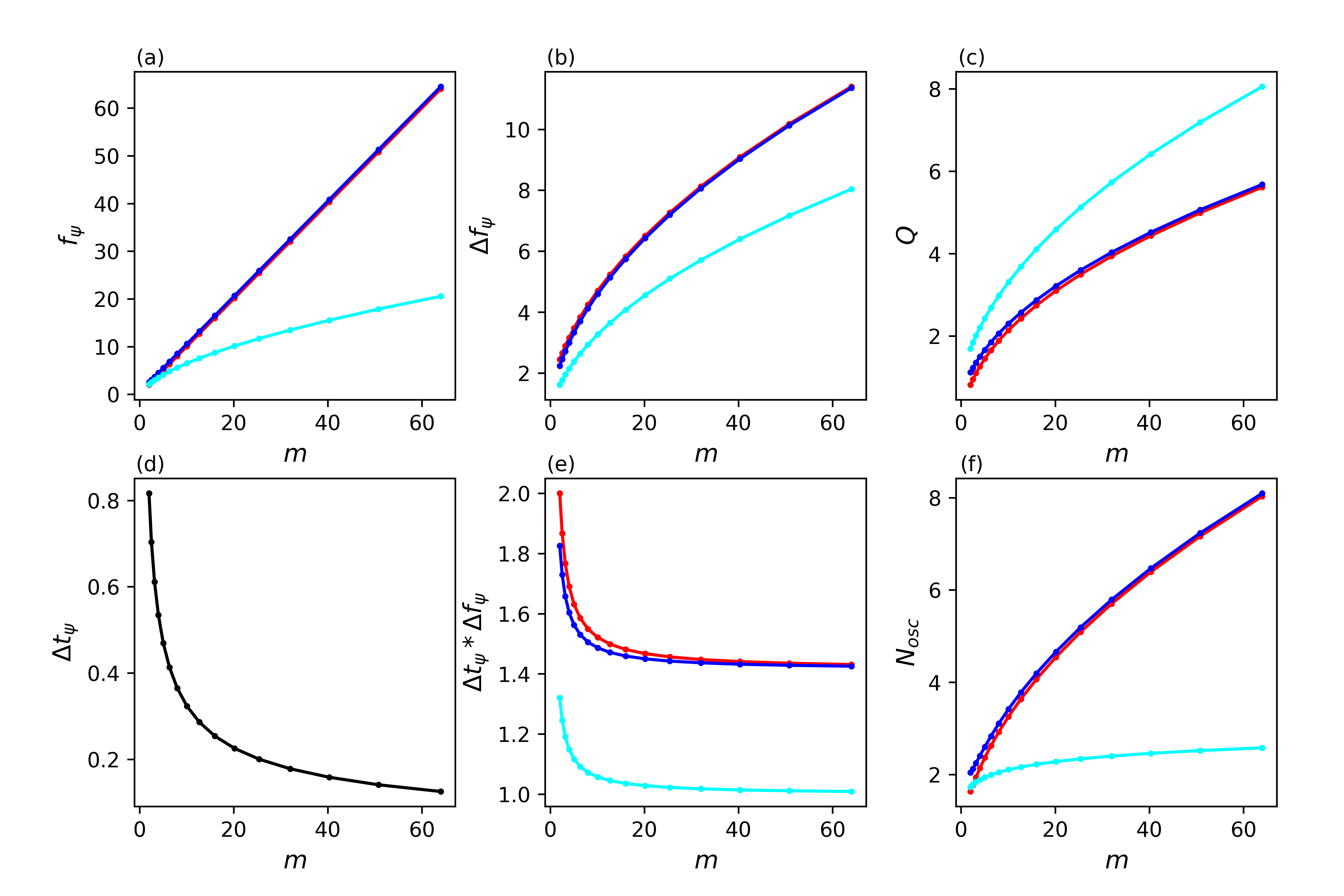}
		\vspace{-0.5em}
		\caption{Representation of the variations of the different Cauchy-Paul wavelet spectral and temporal characteristics with $m$ (and $\tau=1$). (a) Central frequency $f^{[c]}_{\psi_{m}}$. (b) Spectral bandwidth $\Delta f^{[c]}_{\psi_{m}}$. (c) Quality factor $Q$. (d) Temporal bandwidth $\Delta t_{\psi_{m}}$. (d) Temporal and spectral bandwidths product $\Delta t_{\psi_{m}} \Delta f^{[c]}_{\psi}$. (e) Estimated number of temporal oscillations $N_{osc} = \Delta t_{\psi_{m}}f^{[c]}_{\psi_{m}}$. The three centering methods are represented: ($L^{\infty}$ - $c=0$) (red), Centroid ($L^{1}$ - $c=1$) (blue) and Energy ($L^{2}$ - $c=2$) (cyan).}
		\label{fig:Paul_wavelet_evol_char_vs_m}
	\end{figure}
	
	Fig. \ref{fig:Paul_wavelet_evol_char_vs_m} details the variations of the spectral and temporal characteristics of the Cauchy-Paul wavelet as a function of $m$. It compares three centering methods—Peak ($L^{\infty}$), Centroid ($L^{1}$), and Energy ($L^{2}$)—for parameters such as the central frequency $f_{\psi}$, the spectral bandwidth $\Delta f_{\psi}$, the quality factor $Q$, and the temporal bandwidth $\Delta t_{\psi}$. This analysis is fundamental to understanding the optimized time-frequency trade-off within the Cauchy-Paul framework.
	\section{The continuous wavelet transform}
	
	\subsection{Generalities}
	Time-frequency analysis based on the wavelet transform was introduced in the second half of the twentieth century \cite{grossmann_decomposition_1984,carmona_practical_1998} and has since been applied to many scientific domains, including geophysics, fractals, mechanical systems, optics, acoustics, electrophysiology, and cardiology \cite{addison_illustrated_2002}.
	
	The Continuous Wavelet Transform (CWT) of a finite energy signal $x(t) \in L^2(\mathbb{R})$ is defined as its inner product with shifted and scaled copies of an analyzing (mother) wavelet $\psi(t) \in L^1(\mathbb{R}) \cap L^2(\mathbb{R})$ \cite{mallat_wavelet_2009,torresani_analyse_2012}. The wavelet transform of a real signal $x$ is defined using two parameters: a shift parameter $b \in \mathbb{R}$ and a scaling parameter $a \in \mathbb{R}_{+}^{*}$:
	\begin{equation}
		\mathcal{W}_\psi[x](b,a) = \langle x, \psi_{a,b} \rangle = a^{-\frac{1}{p}}\int_{-\infty}^{+\infty} x(t) \psi^*\left( \frac{t-b}{a}\right) dt ,
		\label{eq:waveletdef_t}
	\end{equation}
	where $\psi^*$ is the complex conjugate of the analyzing wavelet $\psi$. The parameter $p$ defines the CWT normalization convention: $p=2$ for an $L^2(\mathbb{R})$ norm (constant energy representation) and $p=1$ for an $L^1(\mathbb{R})$ norm (constant amplitude representation).

	In the frequency domain, the expression of the CWT reads:
	\begin{equation}
		\mathcal{W}_\psi[x](b,a) = a^{1-\frac{1}{p}}\int_{-\infty}^{+\infty} \hat{x}(f) \hat{\psi}^*(af) e^{2i\pi fb} df.
		\label{eq:waveletdef_f}
	\end{equation}
	
	The pseudo-frequency $f_a$ (the frequency of interest) is inversely proportional to the scale $a$ when the center frequency $f^{[c]}_\psi$ is expressed in Hertz:
	\begin{equation}
		f_a = \frac{f^{[c]}_\psi}{a}.
		\label{relation fa et a}
	\end{equation}
	Equation \eqref{relation fa et a} underscores that the scale parameter $a$ is dimensionless.
	Although the peak frequency $f_{\psi}^{[0]}$ is the standard convention for mapping the scale $a$ to the physical frequency $f_a$, the centroid $f_{\psi}^{[1]}$ or the energy frequency $f_{\psi}^{[2]}$ remain valid mathematical alternatives to shift the analytical anchor point.
	
	Table \ref{tab:wavelet_dimensions} presents the dimensional analysis of the CWT: it details the dimensions of $\psi_{m,\tau}$, $\hat{\psi}_{m,\tau}$, $c_{\psi_{m,\tau}}$, and $|\mathcal{W}_{\psi_{m,\tau}}[x](b,a)|$ for both $L^1(\mathbb{R})$ and $L^2(\mathbb{R})$ standard normalizations (see the supplementary material \textbf{S7}  
	for the full derivation).

	\begin{table*}[h]
		\centering
		\caption{Dimensional analysis of the Continuous Wavelet Transform (CWT).}
		\label{tab:wavelet_dimensions}
		\begin{tabular}{llccccl}
			\toprule
			Norm & Scaled & Dimension & Dimension  & Dimension of & Dimension & Primary \\ 
			& wavelet & of $\psi_{m,\tau}$ & $A_{m,\tau}$ or $\hat{\psi}_{m,\tau}$  & $|\mathcal{W}_{\psi_{m,\tau}}[x]\left(a,b\right)|$ & of  $c_{\psi_{m,\tau}}$ & use case \\
			\midrule
			$L^1(\mathbb{R})$ & $\frac{1}{a} \psi\left(\frac{t-b}{a}\right)$ & $1$ & $T$ & $[x] \cdot T$ & $T^2$ & Peak detection, \\
			norm &  & &  &  & & ampl. tracking \\
			\addlinespace
			$L^2(\mathbb{R})$ & $\frac{1}{\sqrt{a}} \psi\left(\frac{t-b}{a}\right)$  & $T^{-1/2}$ & $T^{1/2}$ & $[x] \cdot T^{1/2}$ & $T$ & Energy analysis, \\
			norm &  &  & & & & theor. consistency \\
			\bottomrule
		\end{tabular}
		\label{Table dimension cwt}
	\end{table*}

	The $L^1(\mathbb{R})$ norm ($p=1$) is preferred over the $L^2(\mathbb{R})$ norm for CWT normalization due to three critical advantages. First, it ensures amplitude conservation: the maximum coefficient $|\mathcal{W}(a,b)|$ of a sinusoid remains directly proportional to its physical amplitude $A$, regardless of scale. Conversely, the $L^2$ norm penalizes high frequencies, making a 100 Hz sinusoid appear artificially smaller than a 10 Hz counterpart of identical amplitude. Second, it yields intuitive physical dimensions: as shown in Table \ref{tab:wavelet_dimensions}, the $L^1$ framework ensures the CWT retains the dimension $[x] \cdot T$. Because the mother wavelet is dimensionless, the transform acts as a direct temporal integral, avoiding the non-intuitive $[x] \cdot T^{1/2}$ unit of the $L^2$ norm. Finally, it enables a scale-independent multi-resolution analysis; unlike energy-centric $L^2$ approaches, the $L^1$ norm does not favor low frequencies, ensuring that transient events are visually comparable and detected with equal sensitivity across the entire spectrum.

	\subsection{Estimation of the instantaneous frequency from absolute value and phase of the CWT for a monocomponent signal}
	
	To formally define the instantaneous frequency $f_i(b)$ in Hz, a real monocomponent signal $x(t)$ is mapped to its complex analytic counterpart $z(t)$ using the Hilbert transform $\mathcal{H}\{x(t)\}$, according to the framework pioneered by Ville~\cite{ville1958theory}:
	\begin{equation}
		z(t) = x(t) + i \mathcal{H}\{x(t)\} = A_x(t) e^{i \Phi_x(t)} \; ,
	\end{equation}
	where $A_x(t)$ and $\Phi_x(t)$ are the instantaneous amplitude and phase, respectively.
	The instantaneous frequency $f_i(t)$ (in Hz) is defined as the time derivative of this phase:
	\begin{equation}
		f_i(t) = \frac{1}{2\pi} \frac{d\Phi_x(t)}{dt}.
	\end{equation}
	A comprehensive review of the physical interpretations and mathematical limits of this definition for non-stationary signals can be found in Boashash \cite{boashash1992estimating}. Physically, it represents the local rotational speed of the phase, making it essential for tracking non-stationary variations such as modulations or chirps. 
	Throughout the following of this manuscript, $f_i(b)$ will represent the intrinsic physical instantaneous frequency of the signal, whereas $f_I(b,a)$ will denote the mathematical estimator derived from the complex wavelet transform. 
	
	\noindent The complex CWT is then expressed as $\mathcal{W}_\psi[x](b, a)=|\mathcal{W}_\psi[x](b, a)|\, e^{i \Phi_{\psi}(b,a)}$, thereby separating its magnitude and phase.
	The phase $\Phi_{\psi}(b,a)$ can be extracted via the imaginary part of the complex logarithm of the transform:
	\begin{equation}
		\Phi_{\psi} (b,a) = \mathfrak{Im} \left[ \ln \left( \mathcal{W}_{\psi}[x](b,a) \right) \right] \;.
		\label{Phi=Im of the complex logarithm of CWT}
	\end{equation}

	The squared modulus of the CWT, $|\mathcal{W}_\psi[x](b, a)|^2$, defines the CWT scalogram~\cite{flandrin_fourier_1992}. This quantity represents the time-scale energy density of the signal while discarding all phase information $\Phi_{\psi}(b,a)$. For frequency-modulated components, such as chirps~\cite{chassandemottin1998stationary}, the scalogram provides an intuitive visual representation of the evolution of the instantaneous frequency $f_i(b)$. In their foundational work, Carmona et al.~\cite{carmona_practical_1998} focused on characterizing these signals by extracting structural lines, commonly referred to as ridges, directly from the time-scale plane.

	The estimation of the instantaneous frequency is divided into two distinct approaches depending on the exploited CWT property.

	The first approach relies on ridge extraction methods, based on the principle that the energy of an analytical frequency-modulated signal concentrates along a specific curve in the time-scale plane, referred to as the ``spectral ridge''. This trajectory $a_r(b)$ is estimated by identifying the local maxima of the CWT modulus with respect to the scale parameter $a$:
	\begin{equation}
		a_r(b) = \underset{a}{\arg \max} \left( |\mathcal{W}_{\psi}[x](b,a)| \right) \;,
	\end{equation}
	which satisfies the necessary condition $\frac{\partial}{\partial a} |\mathcal{W}_{\psi}[x](b,a)| = 0$.

	Once the ridge scale $a_r(b)$ is accurately tracked over time using optimization algorithms, such as simulated annealing or ridge-walking penalization techniques, the physical instantaneous frequency is recovered through the relationship:
	\begin{equation}
		f_I^{(1)}(b) = \frac{f^{[0]}_{\psi}}{a_r(b)} \;,
		\label{instantaneous frequency (1)}
	\end{equation}
	where  $f^{[0]}_{\psi}$ is the central frequency of the mother wavelet.
	
	The CWT magnitude, or scalogram, is favored for its phase invariance, allowing robust tracking of frequency modulations and the separation of overlapping components. However, in noisy environments, stochastic fluctuations fragment ridges and generate false trajectories, requiring complex thresholding or energy concentration techniques to extract physical features accurately \cite{iatsenko_extraction_2016}. A foundational analysis of ridge robustness under these conditions is provided by Carmona et al.~\cite{carmona_practical_1998}, while Chassande-Mottin and Flandrin~\cite{chassandemottin1998stationary} pioneered the use of phase and amplitude along these trajectories to precisely identify instantaneous frequencies and damping ratios.

	Alternatively, the second approach leverages the transform's phase information to overcome the energy spreading induced by the wavelet's intrinsic Heisenberg uncertainty. Rather than tracking a single ridge, methods like classical reassignment utilize the local phase derivative to provide an alternative, non-iterative calculation of the instantaneous frequency:
	\begin{equation}
		f_{I}^{(2)} (b,a) = \frac{1}{2\pi} \frac{\partial \Phi_{\psi}(b,a)}{\partial b} \;,
		\label{frequence instantanee-phase}
	\end{equation}
	which circumvents the noise enhancement typical of finite difference methods applied directly to the signal~\cite{carmona_practical_1998}. As a specialized, invertible variant, the Synchrosqueezing Transform (SST) \cite{jiang_instantaneous_2017} can further sharpen this representation by squeezing energy along the frequency axis based on this phase indicator, creating a highly concentrated profile that preserves time resolution for accurate frequency estimation and signal reconstruction.

	Although evaluating the CWT phase is traditionally straightforward via 
	$\Phi_{\psi}(b,a) = \operatorname{atan2}\left(\mathfrak{Im}\{\mathcal{W}_\psi\}, \mathfrak{Re}\{\mathcal{W}_\psi\}\right)$, 
	its numerical stability is severely compromised by phase wrapping. Because the codomain is 
	restricted to $(-\pi, \pi]$, artificial discontinuities appear across the time-scale plane. 
	Recovering a continuous phase profile requires unwrapping algorithms, which are computationally 
	expensive and highly sensitive to noise in non-stationary regimes.
	
	To bypass both the arctangent operator and the need for phase unwrapping, the phase derivative 
	can be extracted directly via the complex logarithm of the transform. Wherever 
	$\mathcal{W}_{\psi}[x](b,a) \neq 0$, the identity 
	$\frac{\partial}{\partial b} \ln\mathcal{W}_{\psi} = \frac{1}{\mathcal{W}_{\psi}}\frac{\partial \mathcal{W}_{\psi}}{\partial b}$ 
	holds, allowing the time-derivative of the CWT phase to be computed via a simple complex ratio:
	\begin{equation}
		\frac{\partial \Phi_{\psi}(b,a)}{\partial b} = \mathfrak{Im} \left( \frac{\frac{\partial}{\partial b} \mathcal{W}_{\psi}[x](b,a)}{\mathcal{W}_{\psi}[x](b,a)} \right) \;.
		\label{eq:wavelet_rev_b}
	\end{equation}
	Combining \eqref{frequence instantanee-phase} and \eqref{eq:wavelet_rev_b} yields the final explicit expression for the phase-derived instantaneous frequency:
	\begin{equation}
		f_{I}^{(2)} (b,a) = \frac{1}{2\pi} \mathfrak{Im} \left( \frac{\frac{\partial}{\partial b} \mathcal{W}_{\psi}[x](b,a)}{\mathcal{W}_{\psi}[x](b,a)} \right) \;.
		\label{frequence instantanee-phase_fin}
	\end{equation}
	
	Since the mother wavelet is a smooth function, the CWT derivative with respect to the translation parameter $b$ can be performed under the integral sign. Under $L^1$ normalization ($p=1$), the differentiation operator affects only the complex exponential kernel, yielding:
	\begin{equation}
		\frac{\partial}{\partial b} \mathcal{W}_\psi[x](b,a) = \int_{-\infty}^{+\infty} (2i\pi f) \hat{x}(f) \widehat{\psi}^*(af) e^{2i\pi fb} \, df \;.
		\label{eq:deriv_WT_b}
	\end{equation}
	
	This expression yields two physical interpretations. First, differentiating the CWT with respect to $b$ is equivalent to computing the CWT of the signal's time derivative $x'(t)$, as the $(2i\pi f)$ term in the Fourier domain represents temporal differentiation. This property is particularly useful for analyzing seismic accelerometric responses without requiring numerical integration~\cite{argoul_use_2025}. Second, the $(2i\pi f)$ term can be absorbed by the wavelet itself, meaning that differentiation with respect to $b$ is equivalent to analyzing the original signal with a frequency-weighted mother wavelet.
	
	It can be easily shown that:
	\begin{equation}
		(2i\pi f) \widehat{\psi}^{*}(af) = -\frac{1}{a} \widehat{\psi'}^{*}(af) \;,
		\label{eq:prop_deriv_fourier}
	\end{equation}
	and \eqref{eq:deriv_WT_b} becomes:
	\begin{align}
		\frac{\partial}{\partial b} \mathcal{W}_\psi[x](b,a) &= -\frac{1}{a} \int_{-\infty}^{+\infty} \hat{x}(f) \widehat{\psi'}^{*}(af) e^{2i\pi fb} \, df = -\frac{1}{a}\, \mathcal{W}_{\psi'}[x](b,a) \;.
		\label{eq:deriv_WT_derivé_ondelette}
	\end{align}
	The instantaneous frequency in \eqref{frequence instantanee-phase_fin}
	\begin{equation}
		f_{I}^{(2)} (b,a) = -\frac{1}{2\pi\,a} \,\mathfrak{Im} \left( \frac{ \mathcal{W}_{\psi'}[x](b,a)}{\mathcal{W}_{\psi}[x](b,a)} \right) \;.
		\label{frequence instantanee-phase_ter}
	\end{equation}
	As long as the mother wavelet $\psi$ is infinitely differentiable ($C^\infty$) and strictly analytic with a bounded positive frequency support, such as the Cauchy-Paul wavelet, its derivative $\psi'$ naturally retains a zero mean ($\widehat{\psi'}(0) = 0$).
	Consequently, $\psi'$ acts as a valid admissible wavelet, ensuring that the phase-derived frequency estimator remains mathematically exact and robust to noise.

	By expressing the phase derivative as an algebraic ratio between the standard CWT and the CWT computed with the differentiated wavelet $\psi'$, this formulation bypasses the numerical instabilities and unwrapping ambiguities of direct differentiation.
	This approach directly leverages the cross-spectral derivative properties pioneered by Nelson~\cite{nelson_cross_spectral_2001} for high-resolution time-frequency analysis.
	
	\section{The Cauchy-Paul continuous wavelet transform}
	
	The continuous Cauchy-Paul wavelet transform (CCWT) has been successfully applied by Muñoz et al. \cite{munoz_continuous_2005} to X-ray absorption fine structure (XAFS) spectra of complex materials, such as those encountered in Earth sciences. This approach provides a three-dimensional mapping that simultaneously correlates the wave vector with the interatomic distance, thereby significantly enhancing the understanding of local atomic structures.
	
	Consequently, this section presents the CCWT along with the proposed methodological enhancements for tracking instantaneous frequency by exploiting the unique structural properties of the Cauchy-Paul wavelet.
	
	By substituting the frequency-domain definition of the Cauchy-Paul mother wavelet $\widehat{\psi}_{m,\tau}(f)$ from \eqref{def-ondelette Cauchy-Paul domaine frequentiel} into the general CWT formulation \eqref{eq:waveletdef_f}, the transform expands to:
	\begin{align}
		& \mathcal{W}_{\psi_{m,\tau}}[x](b,a) = a^{1-\frac{1}{p}}\int_{0}^{+\infty} \hat{x}(f) \widehat{\psi}_{m,\tau}^*(af) e^{2i\pi fb} \, df \nonumber \\
		&= A_{m,\tau} (a\tau)^{m} a^{1-\frac{1}{p}}\int_{0}^{+\infty} \widehat{x}(f) f^{m} e^{-af\tau} e^{2i\pi fb} \, df \;,
		\label{eq:wavelet_cauchy_expanded}
	\end{align}
	where $A_{m,\tau}$ is the normalization factor dictated by the chosen regime $L^p(\mathbb{R})$ for the mother wavelet (see Table~\ref{tab:cauchy_parameters_dim}). For instance, under the standard $L^1(\mathbb{R})$ normalization ($p=1$), this scaling factor reduces strictly to $A_{m,\tau} = \frac{\tau}{m!}$, and the expanded transform from \eqref{eq:wavelet_cauchy_expanded} simplifies to:
	\begin{equation}
		\mathcal{W}_{\psi_{m,\tau}}[x](b,a) = \frac{\tau}{m!} (a\tau)^{m} \int_{0}^{+\infty} \widehat{x}(f) f^{m} e^{-af\tau} e^{2i\pi fb} \, df \;.
		\label{eq:wavelet_cauchy_l1}
	\end{equation}
	
	\subsection{Derivation of the $f_I^{(3)}$ estimator via intrinsic Cauchy-Paul wavelet properties}
	
	By exploiting the differentiation property in the Fourier domain, the derivative of the Cauchy-Paul mother wavelet can be directly mapped to its higher-order counterpart:
	\begin{equation}
		\widehat{\psi'}_{m,\tau}(f) = \frac{2i\pi}{\tau} \frac{A_{m,\tau}}{A_{m+1,\tau}} \, \widehat{\psi}_{m+1,\tau}(f) \;.
		\label{eq:relation_deriv_cauchy_paul}
	\end{equation}
	Consequently, the CWT of $x(t)$ with the differentiated wavelet $\psi'_{m,\tau}$ yields:
	\begin{align}
		\mathcal{W}_{\psi'_{m,\tau}}[x](b,a) &= \int_{0}^{+\infty} \hat{x}(f) \widehat{\psi'}_{m,\tau}^*(af) e^{2i\pi fb} \, df \nonumber \\
		&= -\frac{2i\pi}{\tau} \frac{A_{m,\tau}}{A_{m+1,\tau}} \, \mathcal{W}_{\psi_{m+1,\tau}}[x](b,a) \;.
	\end{align}
	
	When applied to the Cauchy-Paul family, the general phase-based estimator seamlessly reduces to an algebraic ratio of consecutive orders because of its unique structural properties. The instantaneous frequency formulation expressed in \eqref{frequence instantanee-phase_ter} can therefore be directly derived:
	\begin{align}
		f_{I}^{(3)} (b,a) 
		&= \frac{1}{a\tau} \frac{A_{m,\tau}}{A_{m+1,\tau}} \, \mathfrak{Re} \left( \frac{ \mathcal{W}_{\psi_{m+1,\tau}}[x](b,a)}{\mathcal{W}_{\psi_{m,\tau}}[x](b,a)} \right)\;.
		\label{frequence instantanee-phase_four}
	\end{align}

	Alternatively, \eqref{frequence instantanee-phase_four} can be established by bypassing the wavelet derivative $\psi'$ and directly leveraging the recurrence relation of the Cauchy-Paul family:
	\begin{equation}
		\widehat{\psi}_{m+1}(f) = \left( \frac{A_{m+1}}{A_{m}} f \right) \widehat{\psi}_{m}(f) \;.
		\label{eq:recurrence_cauchy_paul}
	\end{equation}
	Evaluating this identity at the scaled frequency $af$ reduces the phase derivative to a straightforward algebraic ratio of consecutive wavelet orders ($m$ and $m+1$). The complete step-by-step proof is detailed in the supplementary material \textbf{S8}.

	\subsection{Multicomponent signals and cross-component interference}
	
	In realistic applications, such as neurophysiological or structural vibration data, the analyzed signal often exhibits a non-stationary ``multicomponent'' structure. Such a signal can be modeled as a linear combination of $N$ distinct mode components, written as:
	\begin{equation}
		x(t) = \sum_{k=1}^{N} x_k(t) = \sum_{k=1}^{N} A_x^{(k)}(t) e^{i \Phi_x^{(k)}(t)} \;.
		\label{Multicomponent signal}
	\end{equation}
	The "multicomponent" aspect denotes a linear superposition ($\sum$) without direct non-linear coupling (e.g., $x_1(t) \cdot x_2(t)$). However, each individual component $x_k(t)$ can be highly non-sinusoidal or stem from a non-linear underlying system.
	The stationary case corresponds to a constant amplitude $A_x^{(k)}(t)$ and a linear phase $\Phi_x^{(k)}(t) = 2 \pi f^{(k)} t + \phi_x^{(k)}$.
	By virtue of the linearity of the CWT, the total transform is the straightforward sum of the individual CWTs associated with each component:
	\begin{equation}
		\mathcal{W}_{\psi_{m,\tau}}[x](b,a) = \sum_{k=1}^{N} \mathcal{W}_{\psi_{m,\tau}}[x_k](b,a) \;.
	\end{equation}
	When evaluating the algebraic ratio of consecutive orders to estimate the instantaneous frequency of a targeted $k$-th component, the neighboring modes inevitably distort the local time-scale plane. The algebraic ratio expands as follows:
	\begin{align}
		\frac{\mathcal{W}_{\psi_{m+1,\tau}}[x](b,a)}{\mathcal{W}_{\psi_{m,\tau}}[x](b,a)} = \frac{\mathcal{W}_{\psi_{m+1,\tau}}[x_k](b,a) + \displaystyle\sum_{j \neq k} \mathcal{W}_{\psi_{m+1,\tau}}[x_j](b,a)}{\mathcal{W}_{\psi_{m,\tau}}[x_k](b,a) + \displaystyle\sum_{j \neq k} \mathcal{W}_{\psi_{m,\tau}}[x_j](b,a)} \;.
	\end{align}
	By isolating the primary contribution of the mode of interest $x_k$, this ratio can be reformulated to incorporate both inner modulation and outer cross-talk effects:
	\begin{align}
		\frac{\mathcal{W}_{\psi_{m+1,\tau}}[x](b,a)}{\mathcal{W}_{\psi_{m,\tau}}[x](b,a)} 
		= \frac{a \tau f_{i}^{(k)}(b)}{A_{m,\tau}^{(k)}/A_{m+1,\tau}^{(k)}}\left( 1 + \epsilon_{\text{chirp}}^{(k)}(b,a,m) + \epsilon_{\text{interf}}^{(k)}(b,a,m) \right) \nonumber \;,
	\end{align}
	where $f_{i}^{(k)}(b) = \frac{\dot{\Phi}_x^{(k)}(b)}{2\pi}$ represents the ideal physical instantaneous frequency of the $k$-th component. The term $\epsilon_{\text{chirp}}^{(k)}(b,a,m)$ is the intrinsic structural error of the isolated $k$-th component, and $\epsilon_{\text{interf}}^{(k)}(b,a,m)$ represents the ``cross-component interference'' term induced by adjacent modes on the targeted $k$-th mode.
	
	The mathematical derivation of the exponential cross-component interference decay is detailed in the supplementary material \textbf{S9}.
	
	Crucially, unlike the monocomponent chirp case, the cross-talk term $\epsilon_{\text{interf}}^{(k)}(b,a,m)$ is a ``fully complex quantity''. Its real part does not vanish ($\mathfrak{Re}\big(\epsilon_{\text{interf}}^{(k)}\big) \neq 0$), meaning that the real-part operator $\mathfrak{Re}$ implemented in the estimator $f_{I}^{(3,k)}(b,a)$ cannot entirely neutralize this cross-talk bias on an arbitrary scale $a$:
	\begin{equation}
		f_{I}^{(3,k)}(b,a) = f_{i}^{(k)}(b) \left( 1 + \mathfrak{Re}\big(\epsilon_{\text{interf}}^{(k)}(b,a,m)\big) \right) \;.
		\label{freq instantaneous multicomponent}
	\end{equation}
	
	When evaluated along a ridge trajectory, the algebraic estimator $f_{I}^{(3,k)}$ yields the instantaneous frequency strictly along the extracted ridge line:
	\begin{align}
		f_{i}^{(k)}(b) \approx f_{I}^{(3,k)}\big(b, a_{r,k}(b)\big)  = \frac{1}{a_{r,k}(b)\tau} \frac{A_{m,\tau}^{(k)}}{A_{m+1,\tau}^{(k)}} \, \mathfrak{Re} \left( \frac{ \mathcal{W}_{\psi_{m+1,\tau}}[x]\big(b,a_{r,k}(b)\big)}{\mathcal{W}_{\psi_{m,\tau}}[x]\big(b,a_{r,k}(b)\big)} \right) \;.
	\end{align}
	Due to the structural properties of the Cauchy-Paul family, the intercomponent interference error decays asymptotically with respect to the wavelet order $m$:
	\begin{equation}
		f_{I}^{(3,k)}\big(b,a_{r,k}(b)\big) = f_{i}^{(k)}(b) \left( 1 + \mathcal{O}\left(e^{-m \chi_{jk}}\right) \right) \;,
		\label{eq:interference_error_decay}
	\end{equation}
	where $\chi_{jk}= \frac{1}{2} \left( \frac{f_{i}^{(j)}(b) - f_{i}^{(k)}(b)}{f_{i}^{(k)}(b)} \right)^2$ represents the relative spectral distance between the cued mode $k$ and its closest neighboring component $j$.
	
	Choosing a sufficiently high order $m$ increases the quality factor ($Q \propto \sqrt{m}$), thereby sharpening the frequency selectivity of the wavelet slice and ensuring a highly accurate, decoupled tracking of overlapping non-sinusoidal modes.
	
	On the ridge trajectory, where $a_{r,k}(b) = \frac{m}{\tau f_{i}^{(k)}(b)}$, intercomponent interference drops rapidly. This localized evaluation ensures that the dominant mode's energy strictly masks adjacent components, while the exponential decay of the Cauchy-Paul window effectively filters out out-of-band cross-terms.
	
	In the remainder of this article, the Cauchy-Paul mother wavelet and the associated CWT are systematically used under the assumptions of $\tau=1$ and $L^1$ normalization. Under these conditions, the expression of the wavelet in the Fourier domain (from \eqref{def-ondelette Cauchy-Paul domaine frequentiel} simplifies to:
	\begin{equation}
		\hat{\psi}_{m}(f) = \frac{1}{m!} f^m e^{-f} H(f) \;,
		\label{def-ondelette Cauchy-Paul domaine frequentiel_tau=1}
	\end{equation}
	where $H(f)$ is the Heaviside step function. Its corresponding time-domain representation is derived as:
	\begin{equation}
		\psi_{m}(t) = \frac{1}{(1 - i 2\pi t)^{m+1}} \;.
		\label{eq:paul_wavelet_time_tau=1}
	\end{equation}
	
	This specific formulation leads to a normalization amplitude of $A_{m} = \frac{1}{m!}$ and an admissibility coefficient of $c_{\psi_{m}} = \frac{(2m-1)!}{2^{2m} (m!)^2}$. By adopting the peak frequency $f^{[0]}_{\psi_{m}} = m$ as the reference, the associated spectral and temporal descriptors are gathered in Table \ref{tab:cauchy_paul_descriptors}.
	
	\begin{table}[htbp]
		\centering
		\caption{Spectral and temporal descriptors of the Cauchy-Paul mother wavelet ($\tau=1$, $L^1$ normalization).}
		\label{tab:cauchy_paul_descriptors}
		\begin{tabular}{lcc}
			\toprule
			\textbf{Descriptor} & \textbf{Symbol} & \textbf{Expression} \\
			\midrule
			Spectral variance   & $(\sigma_f^{[0]})^2$ & $\dfrac{m+1}{2}$ \\
			[1.5ex]
			Spectral bandwidth  & $\Delta f^{[0]}_{\psi_{m}}$ & $\sqrt{2(m+1)}$ \\
			[1.5ex]
			Quality factor      & $Q^{[0]}$ & $\dfrac{m}{\sqrt{2(m+1)}}$ \\
			[1.5ex]
			Time duration       & $\Delta t_{\psi_{m}}$ & $\sqrt{\dfrac{2}{2m-1}}$ \\
			\bottomrule
		\end{tabular}
	\end{table}
	
	By applying the $L^1$-normalization  from \eqref{eq:wavelet_cauchy_l1} to both the mother wavelet and the resulting transform  ($p=1$), and setting $\tau = 1$, the formulation simplifies to:
	\begin{equation}
		\mathcal{W}_{\psi_m}[x](b,a) = \frac{a^m}{m!} \int_{0}^{+\infty} \hat{x}(f) f^m e^{-af} e^{2i\pi fb} \, \mathrm{d}f \;.
		\label{eq:waveletdef_f_p=1_explicit}
	\end{equation}

	The relationship between the scale $a$ and the pseudo-frequency $f_a$ is uniquely defined by the chosen reference frequency $f^{[0]}_\psi$:
	\begin{equation}
		f_a = \frac{f^{[0]}_\psi}{a} = \frac{m}{a} \;.
		\label{relation fa et a_final}
	\end{equation}
	
	Finally, under these identical conditions ($\tau=1$ and $L^1$ normalization), the ratio of successive normalization constants reduces to $A_{m}/A_{m+1} = m+1$. This recurrence property leads to a computationally efficient and robust expression for the instantaneous frequency given in \eqref{frequence instantanee-phase_four}, successfully bypassing the need for trigonometric functions:
	\begin{equation}
		f_{I}^{(3)} (b,a) = \frac{m+1}{a} \mathfrak{Re} \left( \frac{\mathcal{W}_{\psi_{m+1}}[x](b,a)}{\mathcal{W}_{\psi_m}[x](b,a)} \right) \;.
		\label{eq:inst_freq_final_L1}
	\end{equation}

	\section{Applications and performance of the Cauchy-Paul wavelet transform on different signals} 
	
	To validate the performance, accuracy, and noise robustness of the proposed algebraic framework, the methodology is systematically deployed first on benchmarks of synthetic signals with controlled non-stationary properties, and subsequently applied to real-world electrophysiological EEG recordings.

	\subsection{Synthetic signals} 
	
	To evaluate the estimation of the instantaneous frequency $f_{I}^{(3)}\left(b, a\right)$ of a mono-component signal, we consider several typical signals: the impulse function $x(t) = \delta_0(t)$ at $t=0$, a pure harmonic signal $x(t) = X_0 \cos(2\pi \nu_0\,t + \phi_0)$, where $f_i(t) = \nu_0$, and a linear chirp $x(t) = X_0 \cos(2\pi \nu_0 t + \pi \alpha t^2 + \phi_0)$, where the theoretical frequency is $f_i(t) = \nu_0 + \alpha t$.
	
	\subsubsection{Analysis of a impulse signal}
	
	Under impulsive excitation $x(t) = \delta_0(t)$, the signal spectrum is flat ($\hat{x}(f)=\hat{\delta_0}(f)=1$). The CWT under $L^1$ normalization and $\tau=1$ reduces to:
	\begin{equation}
		\mathcal{W}_{\psi_m}[\delta](b, a) = \frac{a^m}{m!} \int_{0}^{+\infty} f^m e^{-af} e^{2i\pi fb} \, \mathrm{d}f \;.
	\end{equation}
	To evaluate this integral, we apply the substitution of complex variables $u = f(a - 2i\pi b)$, which implies $\mathrm{d}f = \frac{\mathrm{d}u}{a - 2i\pi b}$. Through analytic continuation in the complex plane, the integral resolves directly into the standard Gamma function $\Gamma(m+1) = m!$:
	\begin{align}
		\mathcal{W}_{\psi_m}[\delta_0](b,a) = \frac{a^m}{m!} \frac{1}{(a - 2i\pi b)^{m+1}} \int_{0}^{+\infty} u^m e^{-u} \, \mathrm{d}u 
		= \frac{a^m}{(a - 2i\pi b)^{m+1}} \;.
		\label{CWT psi_m impulse Dirac}
	\end{align}
	
	This expression is recovered with the time expression of the Cauchy-Paul wavelet:
	\begin{equation}
		\mathcal{W}_{\psi_m}[\delta_0](b, a) = \frac{1}{a} \psi_m^*\left(-\frac{b}{a}\right)= \frac{1}{a}\frac{1}{\left(1 - i 2\pi \frac{b}{a}\right)^{m+1}}.
	\end{equation}

	Following the exact same mathematical steps for the consecutive order $m+1$, we obtain:
	\begin{align}
		\mathcal{W}_{\psi_{m+1}}[\delta_0](b,a) &= \frac{a^{m+1}}{(m+1)!} \int_{0}^{+\infty} f^{m+1} e^{-f(a - 2i\pi b)} \, \mathrm{d}f = \frac{a^{m+1}}{(a - 2i\pi b)^{m+2}} \;.   
	\end{align}
	
	The local algebraic frequency estimator $f_I^{(3)}(b,a)$ is governed by the real part of the ratio between these two consecutive wavelet orders:
	\begin{equation}
		\frac{\mathcal{W}_{\psi_{m+1}}[\delta_0](b,a)}{\mathcal{W}_{\psi_m}[\delta_0](b,a)} = \frac{\frac{a^{m+1}}{(a - 2i\pi b)^{m+2}}}{\frac{a^m}{(a - 2i\pi b)^{m+1}}} = \frac{a}{a - 2i\pi b} \;.
	\end{equation}
	
	To isolate the real component, we multiply both the numerator and the denominator by the complex conjugate of the denominator, $(a + 2i\pi b)$:
	\begin{equation}
		\frac{a(a + 2i\pi b)}{a^2 + (2\pi b)^2} = \frac{a^2}{a^2 + 4\pi^2 b^2} + i \frac{2\pi a b}{a^2 + 4\pi^2 b^2} \;.
	\end{equation}
	
	Taking the real part reveals that the instantaneous ratio follows a Cauchy-Lorentz distribution centered at $b=0$:
	\begin{equation}
		\mathfrak{Re} \left( \frac{\mathcal{W}_{\psi_{m+1}}[\delta](b,a)}{\mathcal{W}_{\psi_m}[\delta_0](b,a)} \right) = \frac{a^2}{a^2 + 4\pi^2 b^2} \;.
	\end{equation}
	
	The algebraic frequency estimator for the impulse signal is defined as
	\begin{align}
		f_I^{(3)}(b,a) = \frac{m+1}{2\pi a} \, \mathfrak{Re} \left( \frac{\mathcal{W}_{\psi_{m+1}}[\delta_0](b,a)}{\mathcal{W}_{\psi_m}[\delta](b,a)} \right) 
		= \frac{(m+1)a}{2\pi (a^2 + 4\pi^2 b^2)} \;.
		\label{frequency estimator for the impulse signal}
	\end{align}
	The shape of this equation with respect to time $b$ is a bell-shaped curve known as a Cauchy-Lorentz distribution. Before the impulse ($b \to -\infty$), the instantaneous frequency tends toward $0$. At the exact moment of the impulse ($b = 0$), substituting $b = 0$ into the formula yields the maximum maximum value: $f_I^{(3)}(0,a) = \frac{(m+1)a}{2\pi a^2} = \frac{m+1}{2\pi a}$. After the impulse ($b \to +\infty$), the instantaneous frequency decays back towards $0$.

	\subsubsection{Analysis of a single-component bounded harmonic signal}
	
	For a pure sinusoidal signal $x(t) = X_0 \sin(2\pi \nu_0 t + \phi_0)$, the Fourier transform is given by 
	
	\noindent $\widehat{x}(f) = \frac{X_0}{2i} \left[ e^{i\phi_0}\delta(f-\nu_0) - e^{-i\phi_0}\delta(f+\nu_0) \right]$. Substituting this expression into \eqref{eq:waveletdef_f_p=1_explicit} and leveraging the sifting property of the Dirac delta function for positive frequencies ($\nu_0 > 0$) directly yields:
	\begin{equation}
		\mathcal{W}_{\psi_m}[x](b, a) = \frac{X_0}{2i} \frac{(a\nu_0)^m}{m!} e^{-a\nu_0} e^{i(2\pi \nu_0 b + \phi_0)} \;.
		\label{eq:cwt_pure_sine}
	\end{equation}
	
	Both previously described methods for estimating the instantaneous frequency \eqref{instantaneous frequency (1)} and \eqref{eq:inst_freq_final_L1} accurately retrieve $\nu_0$.
	
	For the first approach, the spectral ridge method maximizes the wavelet modulus $|\mathcal{W}_{\psi_m}[x](b,a)|$ with respect to the scale $a$. The ridge trajectory $a_r(b)$ is thus defined by:
	\begin{equation}
		\frac{\partial}{\partial a} \left( a^m e^{-a\nu_0} \right) = 0 \implies a_r(b) = \frac{m}{\nu_0} \;.
	\end{equation}
	Consequently, this ridge-based estimator yields the true invariant frequency:
	\begin{equation}
		f_I^{(1)}(b) = \frac{f^{[0]}_{\psi_m}}{a_r(b)} = \frac{m}{m / \nu_0} = \nu_0 \;.
	\end{equation}

	Meanwhile, for the second approach, the algebraic order-ratio method in \eqref{eq:inst_freq_final_L1} evaluates the $L^1$-normalized CWT at consecutive orders $m$ and $m+1$, as described in  \eqref{eq:cwt_pure_sine}. 
	
	This ratio cancels out all phase and amplitude terms, leading to a purely real expression:
	\begin{align}
		\frac{\mathcal{W}_{\psi_{m+1}}[x](b,a)}{\mathcal{W}_{\psi_m}[x](b,a)} =\frac{a\nu_0}{m+1} \implies 
		f_{I}^{(3)}(b,a) = \frac{m+1}{a} \mathfrak{Re}\left[ \frac{\mathcal{W}_{\psi_{m+1}}[x](b,a)}{\mathcal{W}_{\psi_m}[x](b,a)} \right] = \nu_0 \;.
		\label{eq:methode_rapport}
	\end{align}
	Consequently, while traditional approaches require localization of specific trajectories in the time-frequency plane, the algebraic ratio avoids any numerical optimization and remains valid on any arbitrary scale $a$.
	
\begin{figure}[]
		\centering
		\includegraphics[width=0.48\textwidth]{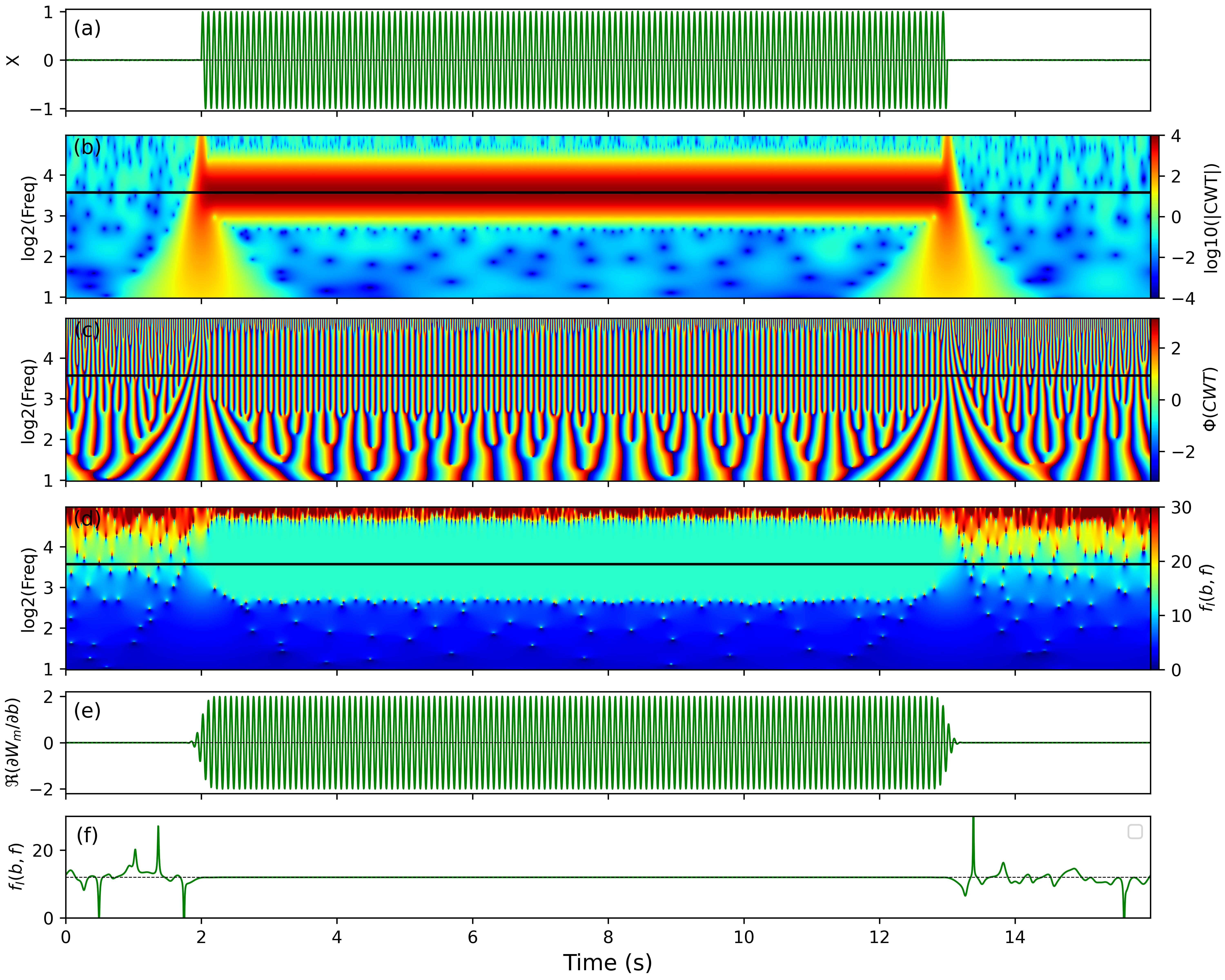}
		\includegraphics[width=0.48\textwidth]{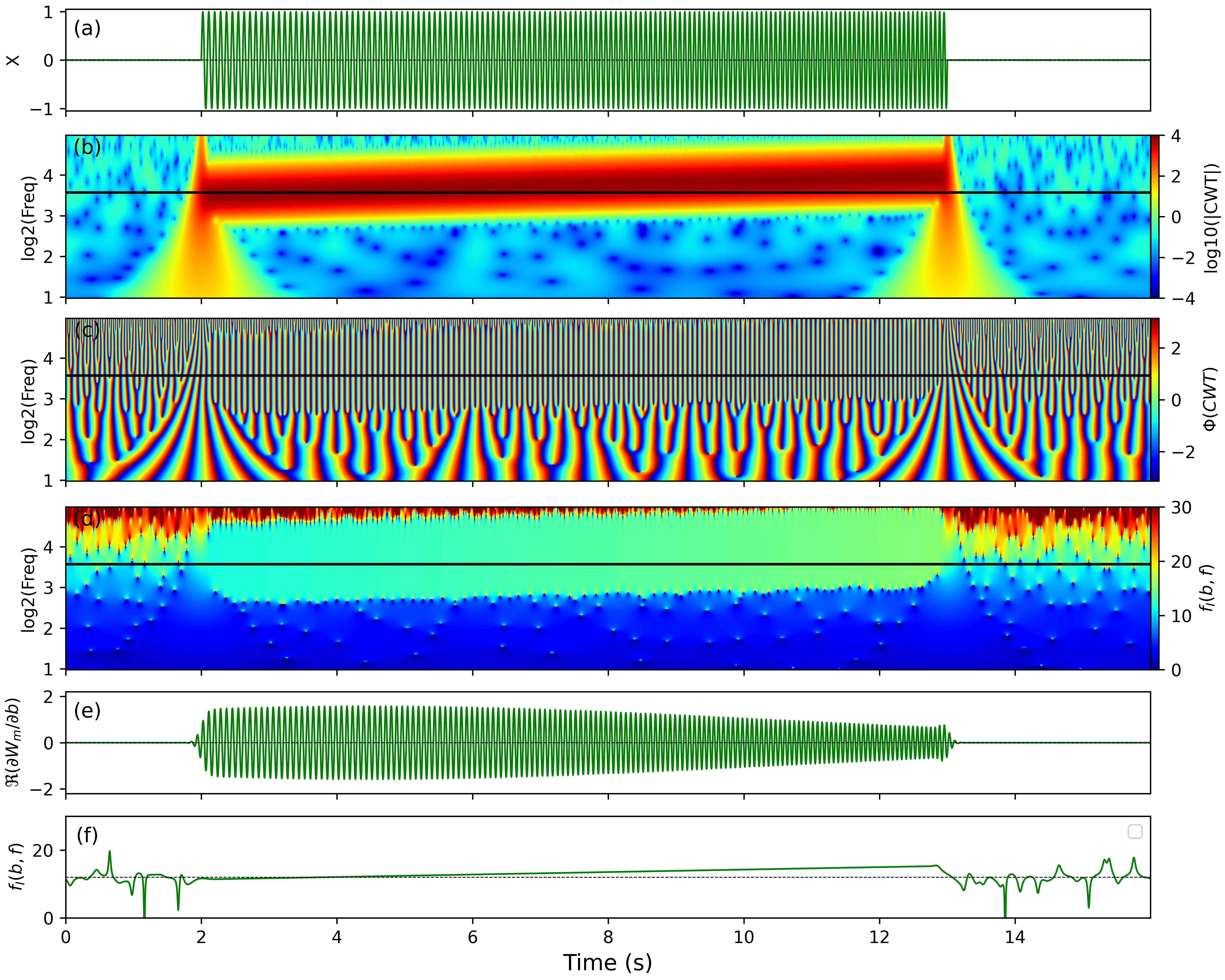}
		\vspace{-0.5em}
		\caption{Time-frequency analysis and direct frequency tracking of a single-component bounded harmonic signal (left) and single-component bounded linear chirp signal (right) using the Cauchy-Paul CWT with $m=32$ and $\tau=1$. (a) Time-domain representation of a bounded $12\text{ Hz}$ harmonic signal $x(t)$. (b) Scalogram displaying $\log_{10}(|\mathcal{W}_{\psi_m}[x]|)$. (c) Wavelet phase distribution $\Phi(\mathcal{W}_{\psi_m}[x])$. (d) Local instantaneous frequency distribution $f_I(b, f)$. (e) Real part of the wavelet coefficient derivative $\Re(\partial \mathcal{W}_{\psi_m}[x]/ \partial b)$ at $\nu^{\top}=12$ Hz. (f) Instantaneous frequency $f_I (b,\nu^{\top})$ with $\nu^{\top}=12$ Hz.}
		\label{cosine_model_mCauchy_32_conphi_phider}
	\end{figure}

	To illustrate and validate the performance of the algebraic instantaneous frequency estimator within the Cauchy-Paul framework, a numerical simulation was conducted using a synthetic single-component signal with abrupt time boundaries, analyzed at a wavelet order of $m = 32$. The results are presented in Fig. \ref{cosine_model_mCauchy_32_conphi_phider}.
	
	\noindent 
	
	The panel \ref{cosine_model_mCauchy_32_conphi_phider} (a) (left column) shows the raw signal in the time-domain $x(t)$, a harmonic oscillation $12\text{ Hz}$ active only within $t \in [2, 13]\text{ s}$: $x(t)= \sin(24\pi t) \mathbb{I}_{[2, 13]}(t) $  where $\mathbb{I}_{[2, 13]}(t)$ denotes the indicator function of the time interval $[2, 13]$. Its CWT log-magnitude scalogram (b) displays a flat energy ridge at $\log_2(\text{Freq}) \approx 3.58$ ($12\text{ Hz}$), bounded by hyperbola-like edge effects from the cone of influence at the time boundaries. This structure matches the regular, parallel phase wrapping shown in the phase plane (c). The estimator performance is detailed in panels (d)–(f). The estimated instantaneous frequency distribution (d) exhibits a stable $12\text{ Hz}$ plateau across the active window, corresponding to the clean oscillatory packet of $\mathfrak{Re}(\partial\mathcal{W}_m/\partial b)$ in panel (e). Finally, the extracted 1D ridge slice (f) confirms that the tracked frequency (solid green) perfectly matches the $12\text{ Hz}$ theoretical baseline (dashed line). The outer fluctuations represent typical Gibbs-like truncation artifacts, validating the high accuracy and selectivity of the algebraic Cauchy-Paul framework away from the boundaries.

	\subsubsection{Analysis of a single-component bounded linear chirp}
	
	For a linear chirp characterized by a quadratic phase $\phi(t) = \phi_0 + 2\pi \nu_0 t + \pi \alpha t^2$, the instantaneous physical frequency is defined as $f_i(b) = \nu_0 + \alpha b$, where the chirp rate is given by $\alpha = \frac{\ddot{\phi}(b)}{2\pi}$. Applying a second-order stationary phase approximation to the CWT around the complex stationary point $f_s = f_i(b) + i \frac{a \tau \alpha}{2\pi}$ yields:
	\begin{align}
		\mathcal{W}_{\psi_{m,\tau}}[x](b,a) &\approx \left[ \frac{X_0}{2 \cdot m!} (a \tau f_i(b))^m e^{-a \tau f_i(b)} e^{i\phi(b)} \right] \cdot 2 e^{i \frac{(a\tau)^2\alpha}{4\pi}} \left( 1 + i \frac{a \tau \alpha}{2\pi f_i(b)} \right)^m \;.
	\end{align}
	
	When evaluating the algebraic ratio of consecutive orders at the same time-scale point $(b,a)$, the amplitude and phase modulation terms cancel out, isolating the following expression:
	\begin{equation}
		\frac{\mathcal{W}_{\psi_{m+1,\tau}}[x](b,a)}{\mathcal{W}_{\psi_{m,\tau}}[x](b,a)} = \frac{a \tau f_i(b)}{m+1} \left( 1 + \epsilon(b, a, m, \tau) \right) \;,
	\end{equation}
	
	\noindent which, when substituted into the algebraic estimator framework, relates directly to the calculated frequency:
	\begin{equation}
		f_{I}^{(3)}(b,a) = f_i(b) \left( 1 + \mathfrak{Re}\big(\epsilon(b,a,m,\tau)\big) \right)\;,
	\end{equation}
	
	\noindent where the structural error term $\epsilon(b, a, m, \tau)$ is found to be ``purely imaginary'':
	\begin{equation}
		\epsilon(b, a, m, \tau) = i \, \frac{a \tau \alpha}{2\pi f_i(b)} = i \, \frac{a \tau \ddot{\phi}(b)}{4\pi^2 f_i(b)} \;.
	\end{equation}
	
	Crucially, since $\mathfrak{Re}\big(\epsilon(b,a,m,\tau)\big) = 0$, the frequency modulation bias is completely neutralized by the real-part operator. Consequently, the estimated instantaneous frequency simplifies to the exact physical trajectory across the entire time-scale plane:
	\begin{equation}
		f_{I}^{(3)} (b,a) = f_i(b) = \nu_0 + \alpha b \;.
	\end{equation}
	
	This unique mathematical property allows the method to achieve a rigorously exact tracking of the instantaneous frequency, even under strong chirp conditions.
	
	To evaluate the adaptability of the algebraic instantaneous frequency estimator under non-stationary conditions, a numerical simulation is performed on a synthetic linear chirp featuring sharp time boundaries, analyzed with a Cauchy-Paul wavelet order $m = 32$. The results are illustrated in Fig. \ref{cosine_model_mCauchy_32_conphi_phider} (right column).
	
	Panel (a) illustrates the raw bounded linear chirp $x(t) = \sin\left( 24\pi t + \pi \frac{2}{11} t^2 \right) \mathbb{I}_{[2, 13]}(t)$, where $\mathbb{I}_{[2, 13]}(t)$ denotes the indicator function of the active time interval. Its nominal instantaneous frequency follows the linear ramp: $f_I(b) = \frac{117}{11} + \frac{2}{11}b$ within the window $t \in [2, 13]$~s. Panel (b) displays the log-magnitude CWT scalogram where the main energy ridge accommodates this linear frequency sweep, tilting upward as time progresses, while the characteristic curved, hyperbola-like diffusion patterns of the cone of influence mark the strong edge effects at the sudden turn-on and turn-off boundaries. This time-scale topology is matched by the phase plane in panel (c), which details the phase wrapping between $-\pi$ and $\pi$ and exhibits a progressive acceleration of the phase patterns beneath the horizontal reference ridge.
	
	The estimator performance and its 1D ridge extractions are highlighted in the subsequent panels. Panel (d) displays the full time-frequency distribution of the estimated instantaneous frequency $f_I(b, f)$, which successfully captures the non-stationary chirp progression across the active window as an ascending, stable tracking zone. Below, panel (e) isolates the 1D cross-section of the real part of the cross-order structural field, $\mathfrak{Re}(\partial\mathcal{W}_m/\partial b)$, along the $12$~Hz frequency line, exhibiting a clear amplitude-modulated envelope that reflects the frequency-induced scale shift. Finally, panel (f) demonstrates the estimator’s tracking accuracy: within $t \in [2, 13]\text{ s}$, the extracted trajectory $f_I(b,f)$ (solid green curve) perfectly follows the linear ramp, crossing the $12\text{ Hz}$ nominal baseline at the simulation mid-point ($t = 7.5\text{ s}$). The minor edge ripples and sharp fluctuations outside this interval correspond to classical truncation artifacts, validating the robust phase-locking capabilities of the algebraic Cauchy-Paul framework under abrupt time boundaries.

	\begin{figure}[]
		\centering
		\includegraphics[width=0.48\textwidth]{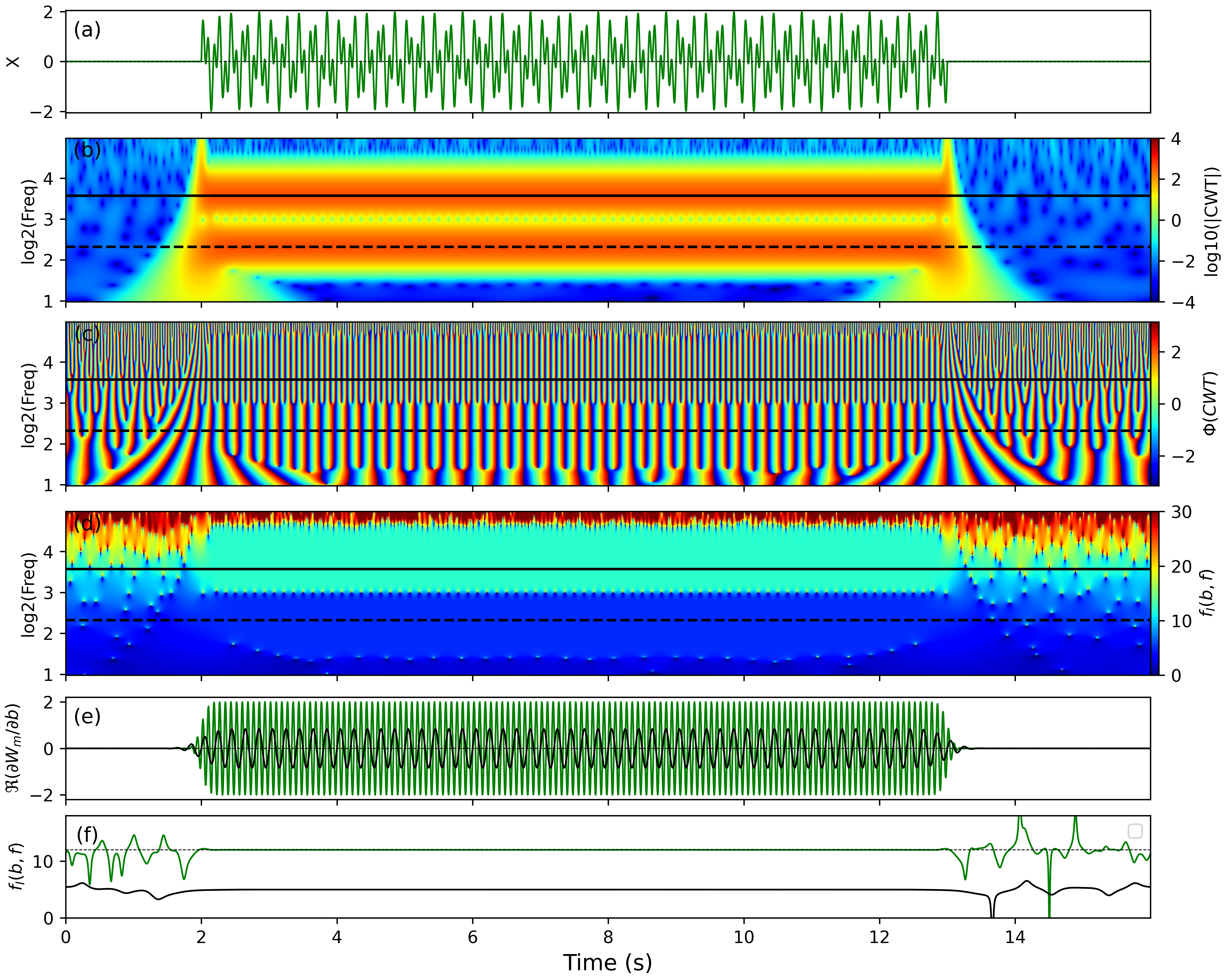}
		\includegraphics[width=0.48\textwidth]{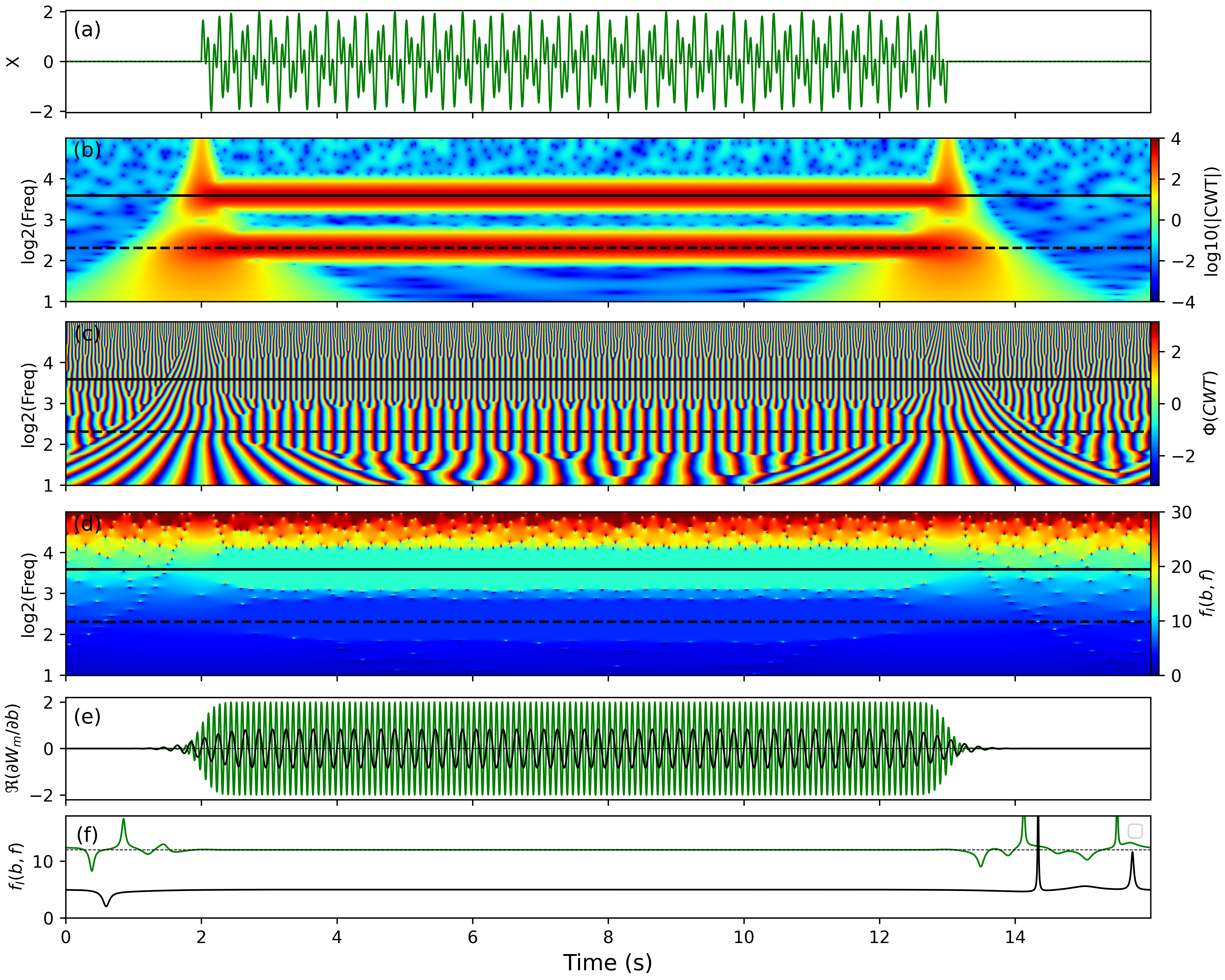}
		\vspace{-0.5em}
		\caption{Time-frequency analysis and direct frequency tracking of a two-component bounded signal (sum of two sines) using the Cauchy-Paul CWT with $\tau=1$,  $m=32$ (left)  and $m=128$ (right). (a) Time history of the analyzed bi-harmonic signal $x(t)$. (b) Scalogram displaying the log-magnitude distribution $\log_{10}|\mathcal{W}_{\psi_m}[x]|$ in the time-frequency plane ($\log_2(f)$ vs. Time), with the black horizontal lines indicating the 12 Hz and the 5 Hz frequency lines. (c) Phase map $\Phi(\mathcal{W}_{\psi_m})$ of the CWT coefficients. (d) Local instantaneous frequency plan $f_I(b, f)$ extracted across the entire time-frequency domain. (e) Cross-section displaying the real part of the cross-order structural field $\mathfrak{Re}(\partial\mathcal{W}_m / \partial b)$ extracted along the $12$ Hz frequency line. (f) Instantaneous frequency $f_I (b,\nu^{\top})$ extracted along the selected frequency lines $\nu^{\top}=12$~Hz (green) and $\nu^{\top}=5$~Hz (black).}
		\label{cosine_sum_mCauchy_32_conphi_phider}
	\end{figure}

	\begin{table*}[h!]
		\centering
		\small
		\caption{Methodological Comparison for a Linear Chirp}
		\begin{tabular}{lll}
			\toprule
			\textbf{Property} & \textbf{Spectral Ridge Method} & \textbf{Algebraic Ratio Method} \\
			\midrule
			Numerical cost & Heavy optimization ($\arg\max_a$) & Direct division of two CWT slices \\
			Scale dependency & Restricted to the ridge $a_r(b) \approx \frac{m}{f_i(b)}$ & Valid at any arbitrary scale $a$ \\
			Chirp Bias & Affected by a systematic ridge shift & \textbf{Rigorously exact} ($f_I(b,a) = f_i(b)$) \\
			\bottomrule
		\end{tabular}
	\end{table*}
	
	\subsubsection{Analysis of a two-component bounded harmonic signal}
	Figure~\ref{cosine_sum_mCauchy_32_conphi_phider}  (left column) illustrates the comprehensive time-frequency decomposition and tracking pipeline for a bounded, bi-component sine signal using a Cauchy-Paul wavelet ($m=32$). 
	The temporal waveform in panel (a) of the left column of Figure~\ref{cosine_sum_mCauchy_32_conphi_phider}  clearly demonstrates the sharp boundaries of the synthetic signal. 
	Panel~(b) shows the resulting scalogram where the two harmonic components are clearly resolved: the primary component, indicated by the solid black horizontal line at $12$~Hz ($\log_2(12) \approx 3.58$), and the secondary, lower-frequency component, marked by the dashed black line at $5$~Hz ($\log_2(5) \approx 2.32$).

	The phase portrait $\Phi(\text{CWT})$ is plotted in panel (c), while panel (d) shows the raw multi-scale instantaneous frequency field $f_i(b, f)$. The stable orange and red plateaus  in panel (b) highlight the area where the phase derivative aligns perfectly with the true signal frequencies. Panel (e) depicts the real part of the spatial derivative $\Re(\partial W_m / \partial b)$, which is critical for computing the phase velocity. Finally, panel (f) presents the instantaneous frequencies $f_{I}^{(3)} (b,f)$ extracted for $f=12$ Hz (green) and for $f=5$ Hz (black). The extracted trajectory matches the expected theoretical frequency with remarkable accuracy, demonstrating the robustness of the method. The minor fluctuations observed at the beginning and the end of the timeline correspond to the expected Cone of Influence (COI) boundary effects due to the finite duration of the bounded signal.
	
	Figure~\ref{cosine_sum_mCauchy_32_conphi_phider}  (right column) shows the same analysis as left column, for a larger value of $m$.

	\begin{figure}[]
		\centering
		\includegraphics[width=0.48\textwidth]{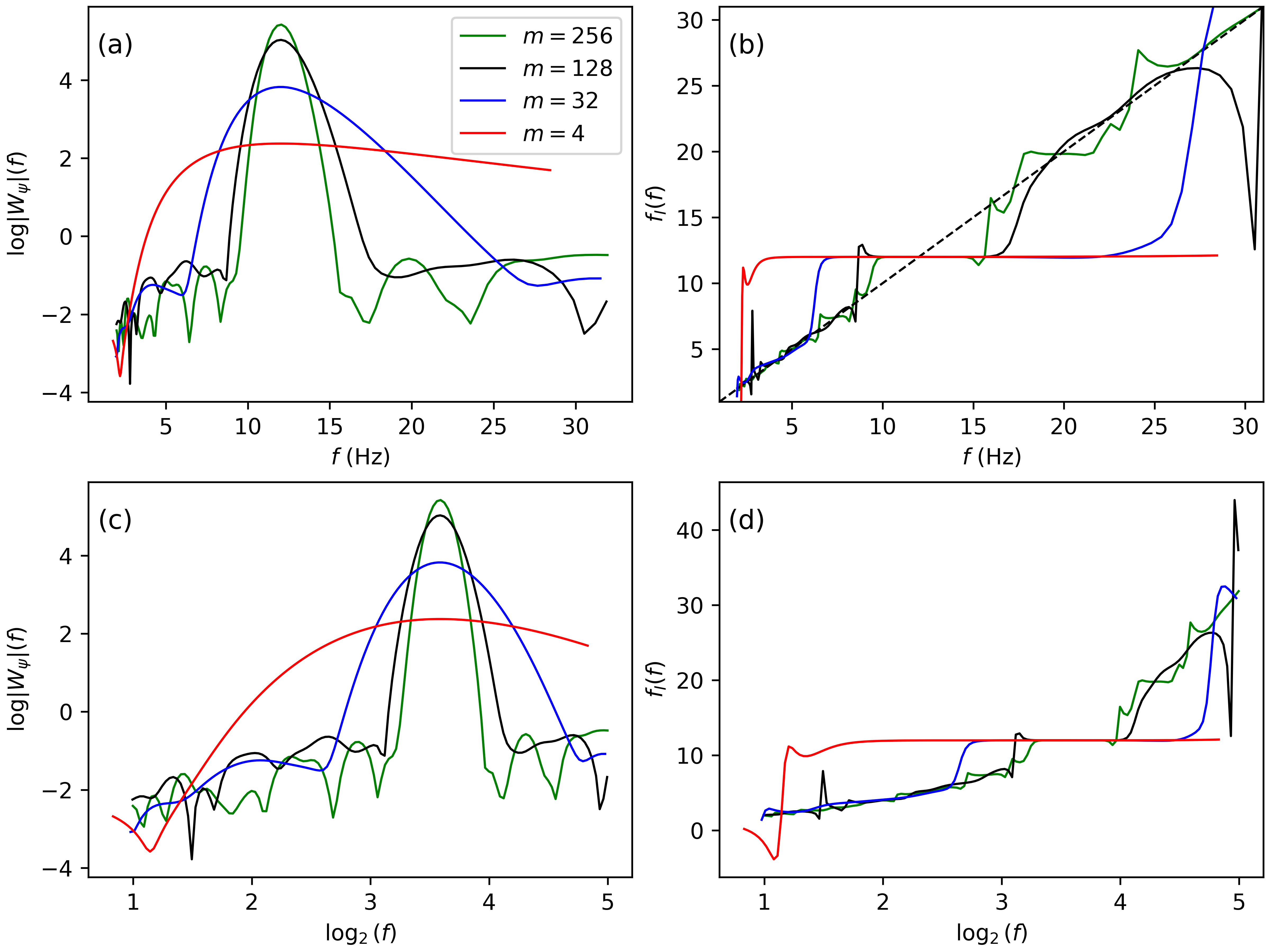} 
		\includegraphics[width=0.48\textwidth]{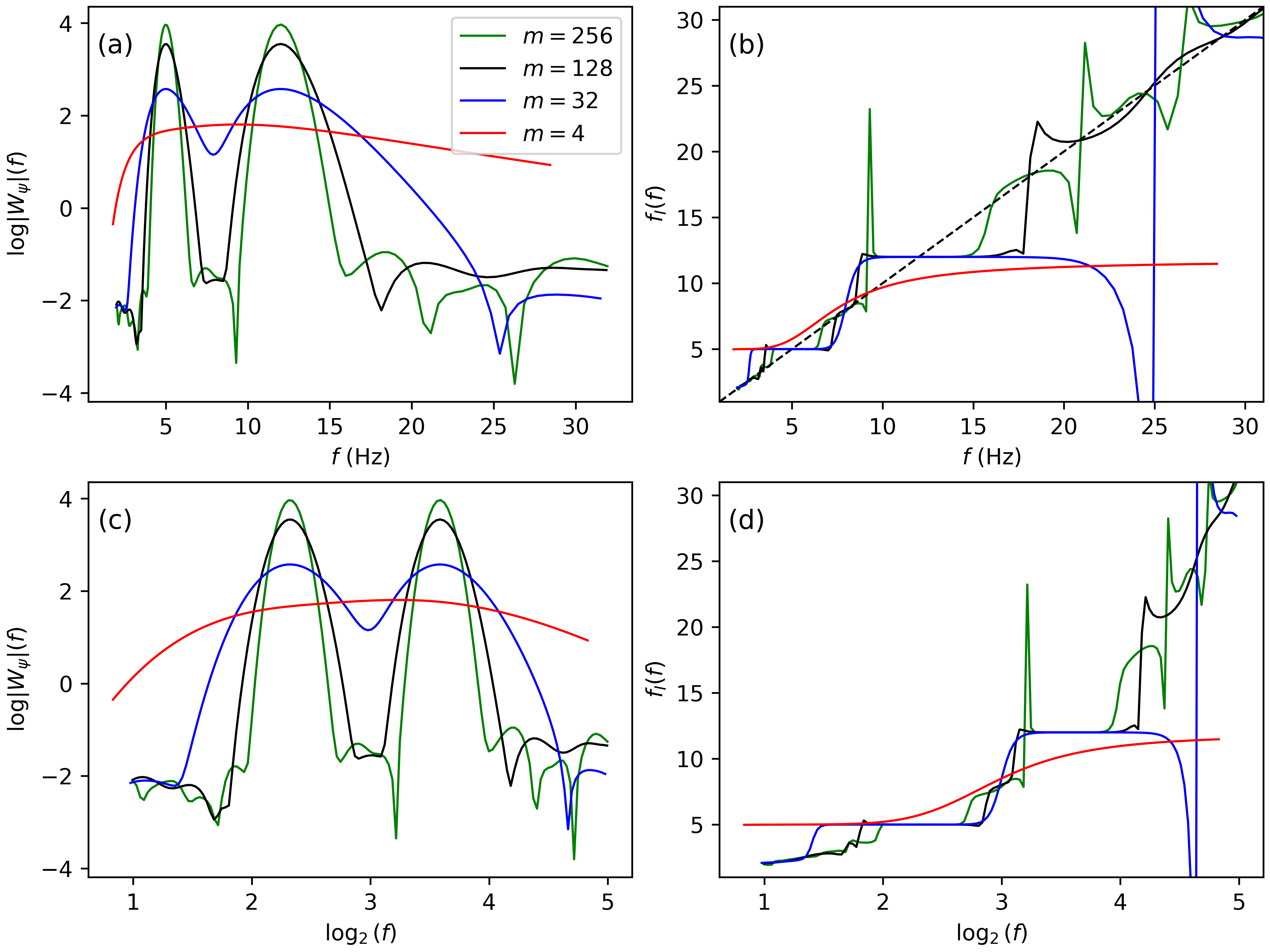}
		\vspace{-0.5em}
		\caption{Frequency profiles of wavelet transform modulus and instantaneous frequency for a sine function (for $b=8\text{~s}$) (left column) and for a sum of two sine functions (right column). (a) $\log_{10} |\mathcal{W}_{\psi_m}[x]|(f)$. (b) $f_I (f)$. }
		\label{compar_profiles_sinaverage_cosine}
	\end{figure}
	
	\subsubsection{Some remarks on two time-frequency estimators  for single-component and two-component bounded sine signals}
	
	The two wavelet-based estimators previously described : the logarithm of the magnitude of the CWT $\log_{10} |\mathcal{W}_{\psi_m}[x]|$ in \eqref{eq:waveletdef_f_p=1_explicit} and the instantaneous frequency $f_I $ in \eqref{eq:inst_freq_final_L1} are plotted in Fig. \ref{compar_profiles_sinaverage_cosine} for the single-component and two-component bounded sine signals for different values of $m$: $m=4,32,128,256$, according to the linear and logarithmic frequency scale.
	Plotted on a $\log_2(f)$ scale, the modulus profiles reveal the intrinsic $\log_f$-symmetry of the Cauchy-Paul wavelet transform. This structural property allows the estimated instantaneous frequency to reach a stable plateau, maintaining a very high estimation accuracy around the spectral peaks. For both single and multi-component signals, selecting a high wavelet order $m$ is essential to narrow the spectral leakage and isolate components. This ensures that in the near vicinity of the ridge trajectory, the instantaneous frequency reaches its stable plateau, maintaining a very high accuracy despite severe edge-interference spikes away from the peaks.

	\subsubsection{Some remarks on the behavior of the estimator under non-linear chirps}
	For a complex non-linear chirp, the higher-order time-derivatives of the instantaneous frequency do not vanish. Consequently, the structural error term $\epsilon(b,a,m)$ is no longer purely imaginary but develops a residual real part, meaning that the algebraic estimator loses its rigorous exactness across the entire time-scale plane and reintroduces a slight dependency on the scale $a$. However, by selecting a high wavelet order $m$ or evaluating the estimator strictly along the spectral ridge trajectory ($a = a_r$), this residual error vanishes asymptotically, maintaining a high tracking accuracy.
	
	Unlike the classical ridge-detection approach, which requires costly time-scale optimization algorithms and introduces systematic biases under strong frequency modulations, the proposed Cauchy-Paul algebraic ratio method inherently minimizes chirp distortions. For linear sweeps or stationary modes, it achieves a non-iterative, mathematically exact estimation of $f_i(b)$ at any arbitrary scale $a$ with minimal computational expense, while offering robust asymptotic convergence for highly non-linear dynamics.

	From a practical standpoint, this asymptotic behavior highlights a classic time-frequency trade-off governed by the Cauchy-Paul mother wavelet order $m$. Increasing $m$ sharpens the frequency selectivity of the Cauchy-Paul filter (see Figure~\ref{cosine_sum_mCauchy_32_conphi_phider}  (right column)), suppresses the non-linear residual error and enforces scale independence. However, this gain expands the wavelet's temporal support, widening the cone of influence and amplifying edge artifacts near abrupt signal boundaries. Optimizing $m$ thus requires a careful balance between the modulation rate of the chirp and the duration of the active window.

	\subsection{Sleep spindle recordings} 
	
	\subsubsection{Deciphering characteristic sleep spectral signatures}
	
	Neural rhythms are observed in wide range of temporal and spatial scales, with very different oscillatory aspects, from quasi-harmonic (periodic) to highly stochastic depending on the brain and body activity (awake, sleep, rest, movement, etc.). 
	Traditionally, these oscillations have been
	clustered into canonical frequency bands, including delta (1-4 Hz), theta (4-8 Hz), alpha (8-12 Hz), beta (15-30 Hz), gamma (30-90 Hz), and high gamma ($>$50 Hz). Spectral analysis of EEG is built on the assumption that EEG rhythms can be approximated by a combination of sinusoidal oscillators. Actually, neural oscillations are nonsinusoidal in most situations. This characteristic of neural oscillations contains crucial information not only about the electrical dynamics of the considered brain zone but also with distant physiological functions \cite{ivanov_new_2021,taillard_sleep_2021}. 
	
	Sleep spindles are recognizable as bursts of quasi-sinusoidal cycles (frequency range 9-16 Hz) from sleeping mammals \cite{fernandez_sleep_2020,sitnikova_sleep_2009,de_gennaro_sleep_2003}. Their name comes from their characteristic fusiform shape. These spindle oscillations occur in the thalamic reticular nucleus (TRN), among which multiple synchronized clusters form \cite{li_spindle_2024}. These fusiform oscillations are typical to a critical state (neighborhood of a Hopf bifurcation) that emerges as the result of the interplay of synchronization and desynchronization forces. Sleep spindles are more prominent in NREM sleep stages, and more specifically during the N2 stage. The duration and shape of the spindle envelope depend on age, cognitive state, brain disorders, sleep regulatory processes, and can also be influenced by hormonal and metabolic factors \cite{yang_regulation_2025,blanco-duque_oscillatory-quality_2024,taillard_sleep_2021}. Interestingly, Blanco-Duque and colls proposed a metrics called oscillatory-Quality to level up spindle synchronization from their underlying neural networks \cite{blanco-duque_oscillatory-quality_2024}, this spectral metric can be compared with  a temporal metric (timing pattern) independently proposed by Chen et al. \cite{chen_individualized_2025}. These recent approaches suggest that a time-frequency metric could be introduced to match the temporal and spectral variability of sleep spindles. Our approach, based on the Cauchy-Paul wavelet transform, is probably a method for reconciling these two aspects, thanks to the possibility of these analytic wavelet families to adapt to the real shape of the transient behaviors of the sleep spindles. 
	
	\subsubsection{EEG maps and electrode nomenclature}
	Electroencephalography (EEG) is today one of the principal tools used in clinics to capture electrical information from brain activity with high temporal resolution and sensitivity \cite{libenson_practical_2025,schomer_niedermeyers_2018}. This technique has also benefited from recent development of complementary imaging techniques (MRI, CT). 
	
	The EEG recordings analyzed in this study were drawn from the FUSO database \cite{coelho_threshold_2025}, which includes polysomnographic data from healthy volunteers. In this work, we focused on recordings from a 24-year-old healthy female subject, using the F3 electrode to demonstrate our CWT-based time-frequency approach.

	\subsubsection{Sleep spindle detection methods}
	
	Traditional expert-based manual examination is both subjective and time-consuming, resulting in significant inter-rate variability. 
	Although the recent development of deep learning or machine assisted learning algorithms, automatic sleep spindle detection remains a persistent challenge in neurophysiology.  The comprehensive benchmark by Warby et al. \cite{warby_sleep-spindle_2014} highlighted low human expert consensus and underscored the limitations of traditional automated detection.
	For instance, many existing algorithms are limited by requirements for prior sleep staging, multi-channel EEG data, or the invalid assumption of signal stationarity as seen in Fourier-based methods.

	Initially, spindle detection relied on the Short-Time Fourier Transform (STFT) to extract time-frequency features, which however proved its limitation from  fixed Fourier windows \cite{prerau_sleep_2017}. This has prompted the development of non-linear sparse time-frequency representations \cite{parekh_sleep_2014} and advanced wavelet-based sparse optimization techniques \cite{parekh_detection_2015} to isolate the complex dynamics of these oscillations. Alternatively, the continuous wavelet transform (CWT) using the Morlet wavelet was successfully introduced by Zygierewicz et al. \cite{zygierewicz_high_1999} to resolve the fine, transient structure of sleep spindles. Their work demonstrated the clear superiority of multi-resolution analysis over fixed-window Fourier transforms for capturing subtle intra-spindle variations through an adjustable time-frequency resolution.
	By tracking instantaneous spindle frequencies with a Morlet CWT in both epileptic and healthy rats, Sitnikova et al. \cite{sitnikova_time-frequency_2014} (2014) revealed an ascending intra-spindle frequency profile unique to the non-epileptic control group.
	Tsanas and Clifford \cite{tsanas_stage-independent_2015} proposed an objective, single-channel EEG spindle detection method using the CWT with the Morlet wavelet and local weighted smoothing. By calculating the "instantaneous strength" of 11–16 Hz coefficients relative to top-ranked values, their algorithm avoided the temporal blurring of fixed-window approaches. However, whilst their methodology demonstrated high sensitivity and specificity across diverse datasets, it faced some limitations such as  reliance on fixed frequency bands, need for heuristic parameter calibration, restricted spatial analysis due to single-channel use, and higher computational demands compared to traditional filtering.
	
	The inherent ability of the CWT to capture transient intermittency is fundamental; unlike Fourier-based approaches that average spectral content over long temporal windows, the CWT  implementation tracks the time-varying coefficient maps over its translation parameter $b$ to precisely isolate the waxing and waning phases of individual spindles.

	\begin{figure}[]
		\centering
		\includegraphics[width=0.48\textwidth]{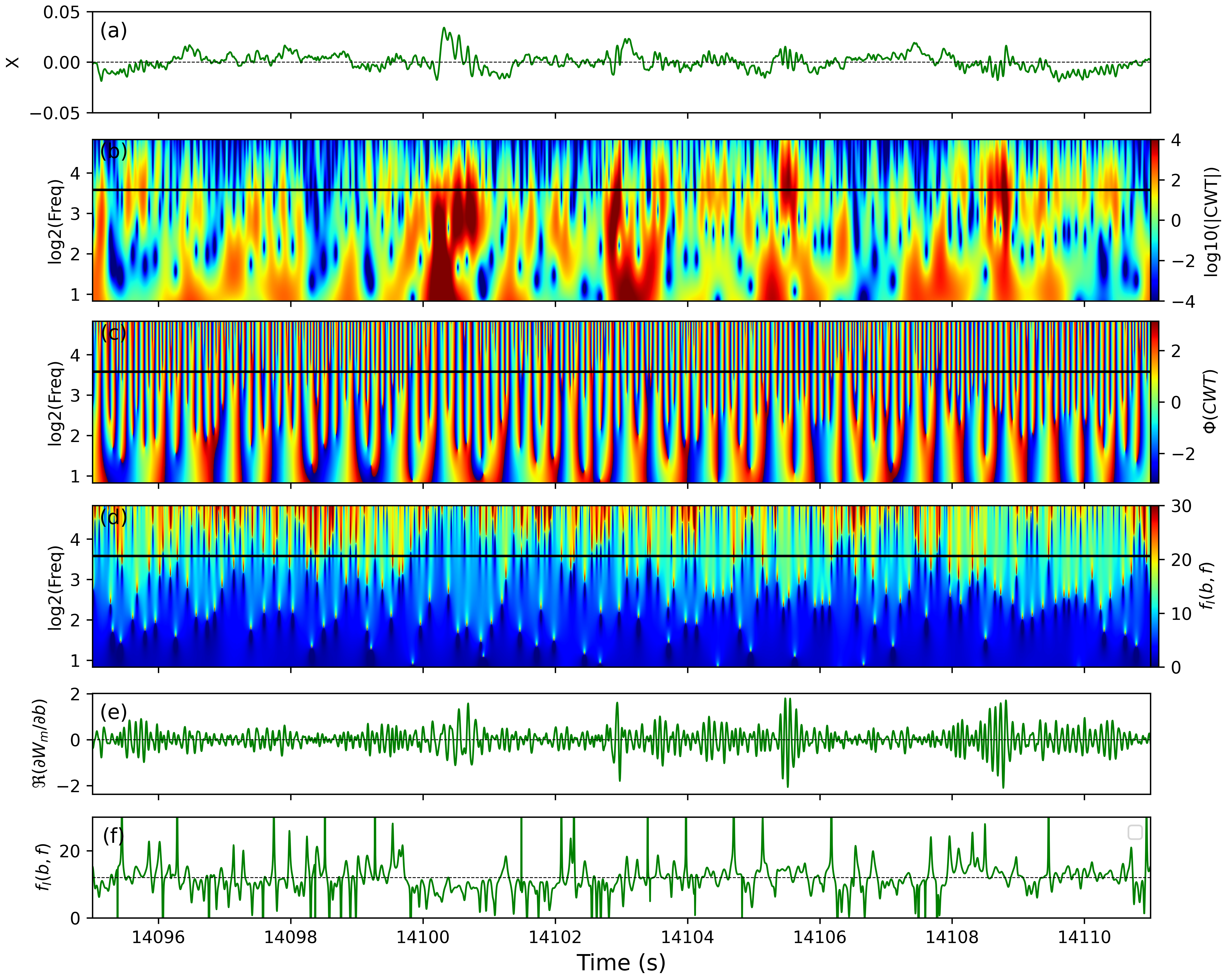}
		\includegraphics[width=0.48\textwidth]{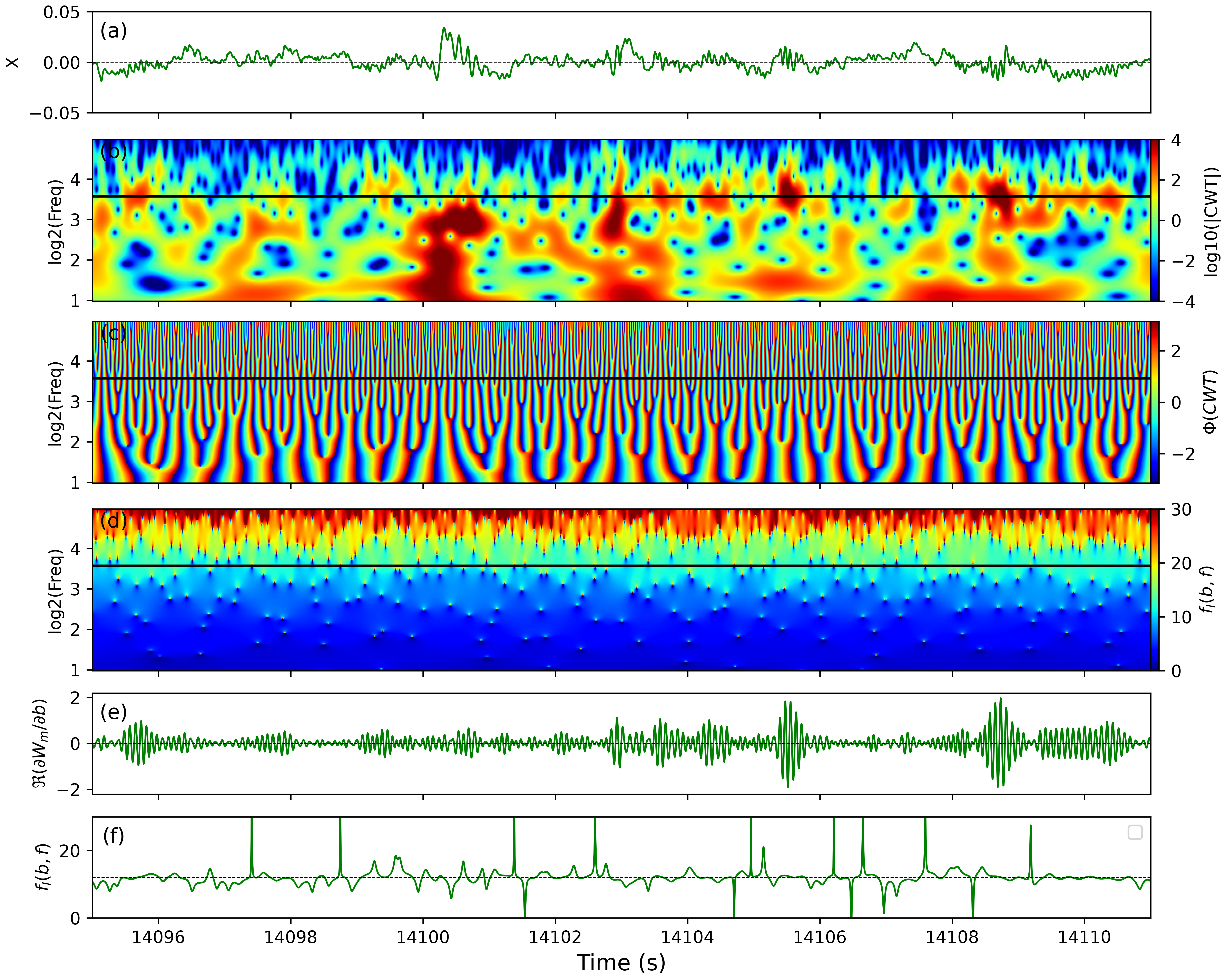}
		\vspace{-0.5em}
		\caption{Time-frequency analysis and instantaneous frequency tracking of a non-stationary sleep EEG recording (Stage N2) using the Cauchy-Paul continuous wavelet transform (CWT) with order $m=4$ (left) and $m=32$ (right). The continuous analysis window spans from $t_{\min} = 14095$\,s to $t_{\max} = 14120$\,s. 
			(a)~Raw EEG signal $x(t)$ displaying characteristic micro-events and transient sleep spindles. 
			(b)~Scalogram representation depicting the log-scaled magnitude of the wavelet coefficients $\log_{10}|\mathcal{W}_{\psi_m}[x]|$ across the $\log_2(f)$ domain. 
			(c)~Phase distribution $\Phi(\mathcal{W}_{\psi_m}[x])$ of the complex wavelet transform coefficients. 
			(d)~Time-frequency distribution of the algebraic phase-based frequency estimator $f_I(b, a)$ mapped into the frequency plane. The solid black horizontal line across panels (b), (c), and (d) denotes the fixed target reference frequency. 
			(e)~Real part of the partial derivative of the wavelet transform with respect to the translation parameter $\mathfrak{Re}(\partial\mathcal{W}_{\psi_m}[x] / \partial b)$, highlighting local structural dynamics. 
			(f)~Extracted instantaneous frequency trajectory $f_I(b, \nu^{\top})$, extracted along the selected frequency line $\nu^{\top}=12$~Hz.}
		\label{tem08_sleephi_N2_Cauchy_m_4_tmin_14095_tmax_14120_conphi_phider_F3}
	\end{figure}

	\subsubsection{Comparative analysis: high temporal ($m=4$) vs. high spectral ($m=32$) resolution}
	
	The practical performance of the Cauchy-Paul continuous wavelet transform (CWT) framework and its associated algebraic phase-based estimator $f_I^{(3)}$ is now illustrated, using a real non-stationary sleep EEG recording (Stage N2) across a continuous 25-second window from $t_{\min} = 14095$\,s to $t_{\max} = 14120$\,s. To evaluate the operational impact of the wavelet structural parameters, two alternative mother wavelet orders are contrasted, namely a low-order framework with $m=4$ shown in Fig.~\ref{tem08_sleephi_N2_Cauchy_m_4_tmin_14095_tmax_14120_conphi_phider_F3} (left) and a high-order framework with $m=32$ shown in Fig.~\ref{tem08_sleephi_N2_Cauchy_m_4_tmin_14095_tmax_14120_conphi_phider_F3} (right).

	The comparison between the low-order ($m=4$) and high-order ($m=32$) Cauchy-Paul CWT frameworks illustrates the fundamental Heisenberg trade-off when tracking non-stationary neural oscillations. At $m=4$, the transform prioritizes temporal resolution, capturing the large-scale envelope and sharp transient boundaries of the raw EEG sleep spindles in panel~(a).
	
	The corresponding log-scalogram $\log_{10}|\mathcal{W}_{\psi_m}[x]|$ in panel~(b) reveals vertically elongated energy clusters that precisely pin down the onset and offset points of these transient micro-events along the $\log_2(f)$ axis, albeit at the expense of tight frequency localization. This rapid tracking is mirrored in the phase map $\Phi(\mathcal{W}_{\psi_m}[x])$ (panel~c) via highly packed, vertical ``phase forks'' or phase singularities. Consequently, the local frequency distribution (panel~d) and its underlying partial derivative $\mathfrak{Re}(\partial\mathcal{W}_{\psi_m}[x] / \partial b)$ (panel~e) capture fast, intracycle fluctuations. Along the multi-component ridges, the extracted instantaneous frequency trajectory $f_I^{(3)}(b, f)$ shown in panel~(f) exhibits a volatile local behavior that fluctuates around the dashed reference line, tracking the immediate structural asymmetry of the spindle while demonstrating the robust phase-tracking capabilities of the $f_I^{(3)}$ framework against non-stationary background noise.
	
	Conversely, implementing a high-order filter ($m=32$) reconfigures the structural geometry of the time-frequency domain by substantially increasing the quality factor ($Q$). While the input signal remains identical, the wavelet coefficients in panel~(b) realign into narrow, horizontal spectral bands, shifting the paradigm entirely toward high frequency resolution. This narrow band-pass filtering behavior yields parallel, highly regular geometric patterns in the phase map (panel~c) and smooths the local frequency distribution (panel~d). The localized phase oscillations $\mathfrak{Re}(\partial\mathcal{W}_{\psi_m}[x] / \partial b)$ in panel~(e) display uniform, harmonic carrier-like profiles. As a result, the ridge-extracted frequency trajectory $f_I^{(3)}(b, f)$ in panel~(f) achieves higher stability, locking onto the dominant spindle carrier frequency in a nearly rectilinear, smooth fashion. By filtering out local noise and microvolt-level fluctuations, the high-order framework effectively averages the spindle's localized energy over a wider temporal interval, smoothing its ``waxing and waning'' envelope at the expense of precise temporal boundary sharpness.
	
	Ultimately, this trade-off highlights the morphological flexibility of the Cauchy-Paul framework governed by the parameter $m$. The configuration $m=4$ preserves temporal integrity and amplitude envelopes, making it optimal for detecting the onset of transient cortical events. Conversely, $m=32$ filters out structural fluctuations to deliver a stable estimation of the carrier frequency. 
	Adjusting $m$ thus allows the framework to be tailored depending on whether the investigation demands sharp time localization or a stabilized frequency characterization under non-stationary noise conditions.

	\begin{figure}[]
		\centering
		\includegraphics[width=0.48\textwidth]{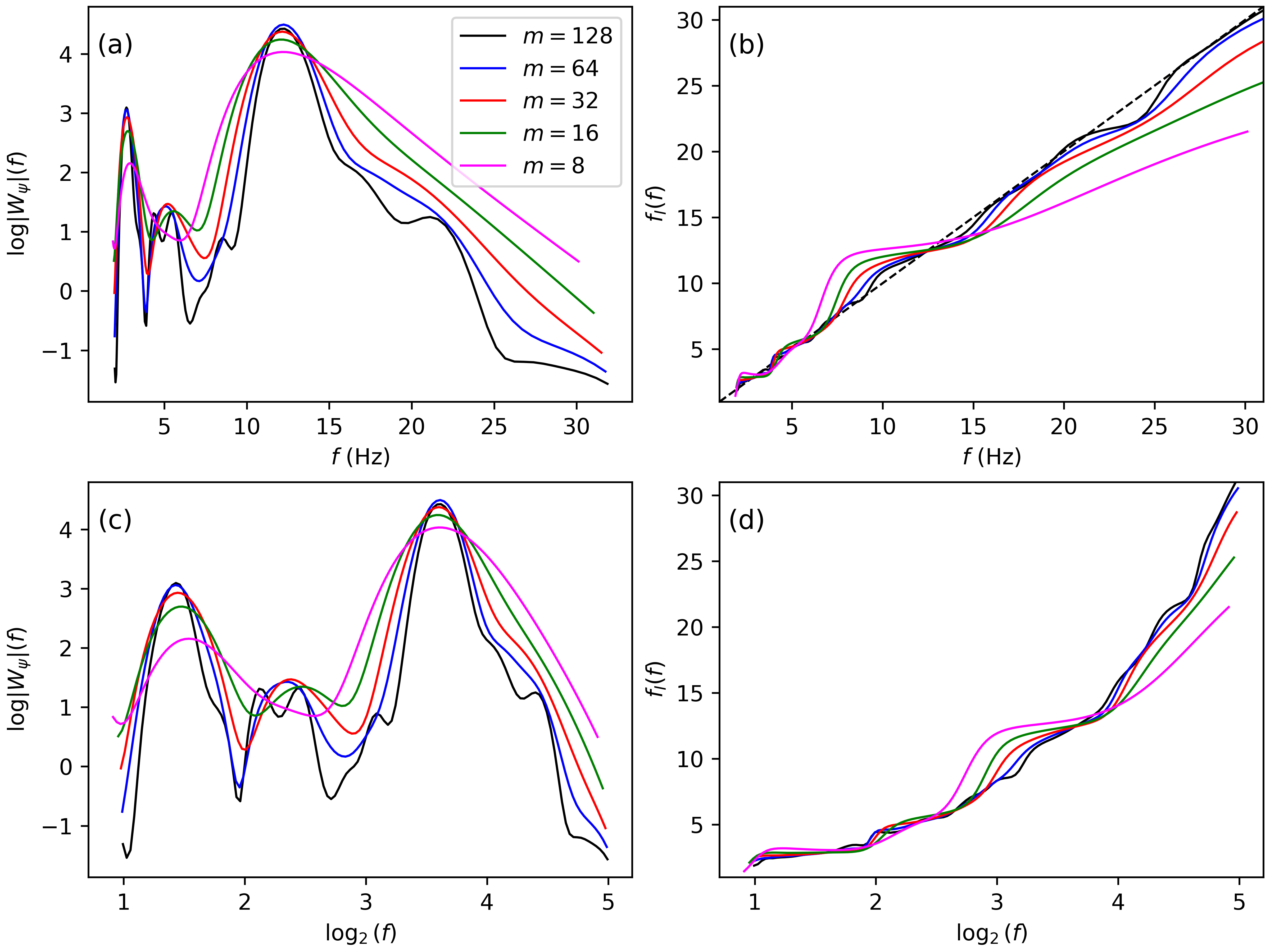}
		\vspace{-0.5em}
		\caption{Frequency profiles of wavelet transform modulus and instantaneous frequency extracted from the EEG signal of Fig. \ref{tem08_sleephi_N2_Cauchy_m_4_tmin_14095_tmax_14120_conphi_phider_F3} (right column) (for $b=14105.55$s). (a) $\log_{10} |\mathcal{W}_{\psi_m}|(f)$. (b) $f_I (f)$. }
		\label{compar_profiles_sinaverage_cosine_more_m}
	\end{figure}

	\section{Discussion - Conclusions}

	The discussion first focuses on the second-order spectral moment frequency $f^{[2]}$, which quantifies the spectral spread of the Cauchy-Paul mother wavelet. While the conventional peak frequency $f^{[0]}$ merely identifies the maximum point of the spectrum, it fails to capture whether the profile is narrow, broad, or skewed. Given that the Cauchy-Paul wavelet is intrinsically asymmetric, featuring a heavy tail toward high frequencies, relying solely on $f^{[0]}$ is insufficient for non-stationary signals. To capture the complete spectral footprint of sleep spindles rather than just their dominant harmonic, the framework leverages the exact spectral centroid $f^{[1]}$ to account for this asymmetric tail, alongside the second-order raw spectral moment $f^{[2]}$ of the Fourier transform $|\widehat{\psi}_{m}(f)|$.
	
	In this architectural characterization, $f^{[2]}$ does not act as a direct instantaneous frequency estimator, but as a critical structural parameter for window calibration. Specifically, it is required to evaluate the spectral variance $\sigma_f^2$ and standard deviation $\sigma_f$ of the filter via the classical moment relation:
	\begin{equation}
		\sigma_f^2 = f^{[2]} - \left(f^{[1]}\right)^2 \;.
	\end{equation}
	
	Within the methodological workflow, this parameter conditions two vital steps. First, regarding ``quality factor $Q$ calibration'', the dimensionless quality factor governing frequency selectivity is defined as $Q = f^{[0]} / \sigma_f$. Since $f^{[2]}$ directly determines $\sigma_f$, it controls the structural tuning of $Q$. Second, for ``Bandwidth Optimization'', the analytical order $m$ must be selected based on this precise spectral envelope to strictly confine the wavelet's effective bandwidth ($B_w = 2\sigma_f$) to the $11$--$16$~Hz sleep spindle frequency band.
	Without accounting for $f^{[2]}$, the asymmetric power-law leakage of the Cauchy-Paul wavelet cannot be quantified, preventing a rigorous isolation of spindle dynamics from out-of-band broadband neural noise.
	
	The discussion next turns to evaluating the local spectral resolution and the geometric stability of the instantaneous frequency plateaus, tracking their behavior across various wavelet orders $m$ within non-stationary sleep EEG signals.
	
	To provide a deeper insight into the local behavior of the analytical framework, the frequency profiles of the wavelet transform modulus and the instantaneous frequency estimator $f_I^{(3)}(f)$ are extracted at a fixed time instant ($b = 14105.55$~s) across multiple wavelet orders ($m = 8, 16, 32, 64, 128$), as depicted in Fig.~\ref{compar_profiles_sinaverage_cosine_more_m}. Panels~(a) and~(c) display the log-scaled modulus $\log_{10}|\mathcal{W}_{\psi_m}|(f)$ plotted against linear frequency and $\log_2(f)$ scales, respectively. For low wavelet orders ($m=8, 16$), the modulus profile appears overly smoothed, largely masking the fine multicomponent substructure of the EEG signal. However, as the structural parameter $m$ increases to $32$, $64$, and $128$, the spectral selectivity increases alongside the filter's quality factor ($Q$), revealing a clear spectroscopic separation of two distinct underlying rhythms. A dominant peak emerges prominently around $12.5$~Hz, corresponding directly to the localized intrinsic signature of the slow sleep spindle, while a secondary, slower peak is isolated in the $2.5$--$3$~Hz delta band. Furthermore, the linear frequency representation in panel~(a) visually demonstrates the inherent positive skewness (spectral asymmetry) of the Cauchy-Paul mother wavelet, showcasing a structural energy spread toward higher frequencies that naturally scales with the frequency-domain argument.
	
	Panels~(b) and~(d) display the corresponding phase-based instantaneous frequency estimator $f_I^{(3)}(f)$ as a function of the filter central frequency. A highly compelling physical phenomenon is observable between $11$~Hz and $13$~Hz, where all tracking curves converge to form a prominent horizontal plateau or stabilization step exactly at $f_I^{(3)} \approx 12.5$~Hz. This horizontal inflection provides unambiguous evidence of a strong, coherent synchronized component within the data. It demonstrates that within this specific bandwidth, regardless of the explicit scale center chosen for the analysis window, the algebraic phase-based estimator successfully corrects the structural mapping law and points directly to the true invariant physical frequency of the sleep spindle. While lower orders yield a smoother, quasi-linear transition, higher orders ($m = 64, 128$) introduce a sharp, S-shaped geometric profile around this plateau, isolating it from neighboring components. A secondary, minor flattening of the estimator's slope is also visible in the low-frequency region near $\log_2(f) \approx 1.5$ ($f \approx 2.8$ Hz) in panel~(d), confirming the presence of the slow delta component independently of the modulus information. At high frequencies ($f > 25$~Hz), where local modal energy vanishes, the higher-order estimator curves asymptotically converge toward the first-bisector identity line ($f_I = f$). This overall behavior demonstrates that the combination of a high-order Cauchy-Paul filter and the $f_I^{(3)}$ estimator can reliably differentiate true physical oscillations, characterized by stable phase velocity plateaus, from background broadband noise, which trivially follows the identity line without forming localized steps.
	
	A third point of discussion concerns the spectral asymmetry of the sleep spindles. This point could serve as a further research comparing the inherent spectral asymmetry of the Cauchy-Paul wavelet with that of the sleep spindle. Biologically, sleep spindles are non-sinusoidal oscillations characterized by rapid onset (waxing) and slower decay (waning) \cite{cole_brain_2017,dimitrov_sleep_2021}, paired with an intra-spindle frequency deceleration (chirp)  \cite{schonwald_quantifying_2011}. This temporal asymmetry distorts the waveform away from a pure sine wave, inherently shifting and stretching spectral energy toward higher frequencies, which results in a positively skewed spectral profile \cite{bullock_bicoherence_1997}.
	Unlike symmetric Gaussian-windowed wavelets (e.g., Morlet), which assume a balanced distribution of energy around the central frequency, the Cauchy-Paul wavelet would exhibit an inherent positive skewness in the frequency domain, defined by its $f^m e^{-f}$ profile. This mathematical asymmetry would mirror the physical and biological reality of sleep spindles, whose spectral densities typically display a right-skewed tail due to their non-sinusoidal waveform shape (sharper edges) and intra-spindle frequency deceleration (chirp). By aligning the asymmetric geometry of the wavelet spectrum with the spindle's actual energy distribution, the Cauchy-Paul framework would act as a spectral matched filter, potentially maximizing the signal-to-noise ratio and ensuring a more precise, unbiased extraction of the instantaneous envelope and phase. Importantly, within the objective to generalize such an analysis to large datasets spanning whole nights, it will be important to implement systematic algorithms to remove artifacts  (transient and possibly large amplitude) \cite{dora_adaptive_2022} and to separate physiological noise from EEGs, to improve the spindle detection with the Cauchy-Paul wavelet.
	
	The Cauchy-Paul wavelet is uniquely suited for the study of sleep spindles due to its mathematical flexibility and inherent analyticity. Its advantages over traditional wavelets, such as the Morlet wavelet, are substantial. First, its perfect analyticity allows for the precise extraction of instantaneous phase and amplitude without the artifacts typically associated with Hilbert transform approximations. Second, the order parameter $m$ enables fine-tuned spectral localization, allowing the detection window to be optimized specifically for the 11--16 Hz spindle frequency band while minimizing background noise. Leveraging its exact centroid frequency $f^{[1]}$ and its second-order spectral moment $f^{[2]}$ provides a much more rigorous physical framework than that of the peak frequency ($f^{[0]}$).
	Finally, its power-law temporal decay ensures superior resolution for detecting spindle onset and offset, effectively preventing the temporal smearing common in fixed-window approaches.
	
	In conclusion, the Cauchy-Paul wavelet provides a highly robust framework for time-frequency analysis, offering superior precision in both temporal localization and spectral characterization of non-stationary electrophysiological events.
	
	\section*{Author Contributions}
	Conceptualization: PA, JT, FA. 
	Data curation: JT,
	Formal analysis: PA, FA,
	Funding acquisition: JT,
	Investigation: PA, JT, FA,
	Methodology: PA, JT, FA,
	Software: FA,
	Supervision: PA, JT, FA,
	Validation: PA, JT, FA,
	Visualization: PA, JT, FA,
	Writing – original draft: PA, FA,
	Writing – review and editing: PA, JT, FA.
	
	\section*{Acknowledgments}
	We thank the Bordeaux University Hospital, Bordeaux University, and
	French National Centre for Scientific Research (CNRS) for their support in the conception and implementation of this study.
	
	\section*{Funding}
	This work was supported by Agence Nationale de la Recherche, Labex BRAIN ANR-10-L ABX-43.
	
	\section*{Declaration of Generative AI in the writing process}
	During the preparation of this work, PA used Google Gemini in order to refine the English language, improve the editorial phrasing. After using this tool, the author reviewed and edited the content as needed and take full responsibility for the scientific integrity and final layout of the manuscript.
	
	\section*{Data Availability Statement}
	The data that support the findings of this study are available on request from FA. The data are not publicly available due to privacy or ethical restrictions.

	\bibliographystyle{cas-model2-names}

	
	\clearpage
	\newpage
	
\end{document}


\maketitle

\thispagestyle{fancy} 

\begin{abstract}
This document contains the supplementary material and appendices associated with the main manuscript.
\end{abstract}

\hrule
\vspace{1em}

\section*{S1. The admissibility coefficient for the Cauchy-Paul mother wavelet}
\label{ap:The admissibility coefficient for the Cauchy-Paul mother wavelet}

The frequency-domain definition of the Cauchy-Paul wavelet is:
\begin{equation}
    \hat{\psi}_{m, \tau}(f) = A_{m, \tau} (\tau f)^m e^{-\tau f} H(f).\nonumber
\end{equation}
where $H(f)$ is the Heaviside step function. The admissibility coefficient is defined by the integral:
\begin{equation}
    c_{\psi_{m,\tau}} = \int_{0}^{\infty} \frac{|\hat{\psi}_{m,\tau}(f)|^2}{f} df \;. \nonumber
\end{equation}
Substituting the definition $\hat{\psi}_{m,\tau}(f) = A_{m,\tau} (f \tau)^m e^{-f \tau}$:
\begin{align*}
    c_{\psi_{m,\tau}} = A_{m,\tau}^2 \int_{0}^{\infty} \frac{(f \tau)^{2m} e^{-2f \tau}}{f} df 
    = A_{m,\tau}^2 \tau^{2m} \int_{0}^{\infty} f^{2m-1} e^{-2\tau f} df\;.
\end{align*}
Using the Gamma function identity $\int_{0}^{\infty} x^{n-1} e^{-ax} dx = \frac{\Gamma(n)}{a^n}$ with $n = 2m$ and $a = 2\tau$:
\begin{equation}
    c_{\psi_{m,\tau}} = |A_{m,\tau}|^2 \tau^{2m} \frac{\Gamma(2m)}{(2\tau)^{2m}} = \frac{A_{m,\tau}^2}{2^{2m}} \Gamma(2m) \;.\nonumber
\end{equation}

\section*{S2. Time-domain derivation of the Cauchy-Paul wavelet}
\label{ap:wavelet_Time-domain Derivation}
Starting from the frequency-domain definition of the Cauchy-Paul mother wavelet:
\begin{equation*}
    \hat{\psi}_{m,\tau}(f) = A_{m,\tau} (f\tau)^m e^{-f\tau} H(f) \; .
\end{equation*}
with $A_{m,\tau}=\frac{\tau}{m!}$ when the $L^1(\mathbb{R})$ norm is chosen and 
$A_{m,\tau}=$ when the $L^2(\mathbb{R})$ norm is preferred.

To find the time-domain expression $\psi_{m,\tau}(t)$, we compute the inverse Fourier transform using the frequency convention:
\begin{equation*}
    \psi_{m,\tau}(t) = \int_{-\infty}^{+\infty} \hat{\psi}_{m,\tau}(f) e^{2i\pi ft} df \;.
\end{equation*}

Since the Heaviside step function $H(f)$ restricts the integration range to $[0, +\infty)$, we have:
\begin{equation*}
    \psi_{m,\tau}(t) =  A_{m,\tau} \int_{0}^{+\infty} (f\tau)^m e^{-f\tau} e^{2i\pi ft} df \;.
\end{equation*}

By grouping the exponential terms, the expression becomes:
\begin{equation*}
    \psi_{m,\tau}(t) =  A_{m,\tau} \tau^{m} \int_{0}^{+\infty} f^m e^{-f(\tau - 2i\pi t)} df \;.
\end{equation*}

We use the standard identity for the Gamma integral:
\begin{equation*}
    \int_{0}^{\infty} x^n e^{-\alpha x} dx = \frac{n!}{\alpha^{n+1}}, \quad \text{for } \text{Re}(\alpha) > 0  \;.
\end{equation*}

By setting $n = m$ and $\alpha = (\tau - 2i\pi t)$, we obtain:
\begin{equation*}
    \psi_{m,\tau}(t) =  A_{m,\tau} \tau^{m} \cdot \frac{m!}{(\tau - 2i\pi t)^{m+1}} \;.
\end{equation*}

When the $L^1(\mathbb{R})$ norm is chosen, $A_{m,\tau}=\frac{\tau}{m!}$ and after simplifying the factorials, we arrive at the final time-domain expression:
\begin{equation*}
    \psi_{m,\tau}(t) = \left( \frac{\tau}{\tau - 2i\pi t} \right)^{m+1} \;.
\end{equation*}

\noindent \textbf{Verification at $t=0$:} Substituting $t=0$ yields the following:
\begin{equation*}
    \psi_{m,\tau}(0) = \left( \frac{\tau}{\tau - 0} \right)^{m+1} = 1 \;.
\end{equation*}
This result is perfectly consistent with the choice of normalization $L^1$ in the frequency domain. Indeed, by the property of the inverse Fourier transform, $\psi(0) = \int_{-\infty}^{+\infty} \hat{\psi}(f) df$. The unit value at the origin confirms that the spectral amplitude $A_{m,\tau}$ was set correctly to ensure $\int_{0}^{\infty} \hat{\psi}_{m,\tau}(f) df = 1$.

When the $L^2(\mathbb{R})$ norm is chosen, we substitute $A_{m,\tau} = \frac{2^m \sqrt{2\tau}}{\sqrt{(2m)!}}$. The expression becomes:
\begin{equation*}
    \psi_{m,\tau}(t) = \frac{2^m \sqrt{2\tau}}{\sqrt{(2m)!}} \cdot \frac{m! \tau^m}{(\tau - 2i\pi t)^{m+1}} \;.
\end{equation*}
By factoring out $\tau^{m+1}$ from the denominator, we obtain the $L^2$-normalized form:
\begin{equation*}
    \psi_{m,\tau}(t) = \frac{2^m m! \sqrt{2}}{\sqrt{(2m)! \tau}} \left( \frac{\tau}{\tau - 2i\pi t} \right)^{m+1} \;.
\end{equation*}
\noindent \textbf{Verification at $t=0$:} At the origin, we have $\psi_{m,\tau}(0) = \frac{2^m m! \sqrt{2}}{\sqrt{(2m)! \tau}}$, which exactly matches the result obtained from the energy-preservation requirement $\|\hat{\psi}_{m,\tau}\|_{L^2}=1$. This value carries the dimension $T^{-1/2}$, ensuring that the total energy of the wavelet remains unitary.

\section*{S3. Expression of the Cauchy-Paul mother wavelet parameters according to the chosen norm}
\label{ap:Expression of the Cauchy-Paul mother wavelet parameters according to the chosen norm}

The computation of the $A_{m,\tau}$ and $c_{\psi_{m,\tau}}$ parameters for the Cauchy-Paul mother wavelet is detailed below, according to the two chosen normalization conventions: the $L^1(\mathbb{R})$ norm and the $L^2(\mathbb{R})$ norm.

\noindent \textbf{If the $L^1(\mathbb{R})$ norm is chosen:}
$|\hat{\psi}_{m,\tau}|_{L^1} = \int_{0}^{\infty} |\hat{\psi}_{m,\tau}(f)| \d f = 1$ and  $[A_{m,\tau}] \cdot [\d f] = 1 \implies [A_{m,\tau}] \cdot T^{-1} = 1$ the dimension of $A_{m,\tau}$ and of $\hat{\psi}_{m,\tau}$ carries the dimension of time (e.g., seconds) is:  $\mathbf{[A_{m,\tau}] = \left[\hat{\psi}_{m,\tau}\right] = T}$.

The dimension of $\psi_{m,\tau}$ can be deduced from this: $[\psi_{m,\tau}]= \frac{[\hat{\psi}_{m,\tau}]}{T}=\frac{T}{T}=1$.

Moreover 
$ \int_{0}^{\infty} |\hat{\psi}_{m,\tau}(f)| \d f = \int_{0}^{\infty} A_{m,\tau} \, (f \tau)^m e^{-f \tau} \d f = 1$, using the substitution $u = f\tau$, we obtain:
\begin{equation*}
\frac{A_{m,\tau}}{\tau} \int_{0}^{\infty} u^{m} e^{-u} du = 1 \implies A_{m,\tau} = \frac{\tau}{\Gamma(m+1)}=\frac{\tau}{m!} \;.
\end{equation*}

We check that the dimension of $A_{m,\tau}$ is time and the admissibility constant $c_{\psi_{m,\tau}}$ is then:
\begin{equation*}
c_{\psi_{m,\tau}} = \frac{A_{m,\tau}^2}{2^{2m}} \Gamma(2m) = \frac{\tau^2}{\Gamma(m+1)^2} \frac{\Gamma(2m)}{2^{2m}}=\frac{\tau^2 (2m-1)!}{2^{2m} (m!)^2} \;.
\end{equation*}
Physical Dimension: $[c_{\psi_{m,\tau}}] = T^2$ (Time squared).
We check that: $\left[c_{\psi_{m,\tau}}\right]\,=\,\left[\left|\hat{\psi}_{m,\tau}\right| \right]^{2}$.


\noindent \textbf{If the $L^2(\mathbb{R})$ norm is chosen:} We impose the condition $\|\psi_{m,\tau}\|_{L^2(\mathbb{R})} = \|\hat{\psi}_{m,\tau}\|_{L^2(\mathbb{R})} = 1$. This choice implies the following dimensional relations:
\begin{equation*}
    [A_{m,\tau}]^2 \cdot T^{-1} = 1 \implies [A_{m,\tau}] = [\hat{\psi}_{m,\tau}] = T^{1/2} \;.
\end{equation*}
The dimension of the wavelet $\psi_{m,\tau}$ in the time domain is then:
\begin{equation*}
    [\psi_{m,\tau}] = \frac{[\hat{\psi}_{m,\tau}]}{T} = \frac{T^{1/2}}{T} = T^{-1/2} \;.
\end{equation*}

To ensure unitary energy, the Fourier transform must satisfy:
\begin{equation*}
    \|\hat{\psi}_{m,\tau}\|_{L^2}^2 = \int_{0}^{\infty} |A_{m,\tau}|^2 (f\tau)^{2m} e^{-2f\tau} df = 1
\end{equation*}
Using the substitution $u = 2f\tau$ (which implies $df = \frac{du}{2\tau}$), we obtain:
\begin{equation*}
    \frac{A_{m,\tau}^2}{2^{2m+1}\tau} \int_{0}^{\infty} u^{2m} e^{-u} du = 1\implies A_{m,\tau}^2 = \frac{2^{2m+1}\tau}{\Gamma(2m+1)} = \frac{2^{2m+1}\tau}{(2m)!} \;.
\end{equation*}
Since $A_{m,\tau}$ is a real positive constant, it follows that:
\begin{equation*}
    A_{m,\tau} = \sqrt{\frac{2^{2m+1}\tau}{(2m)!}} = \frac{2^m \sqrt{2\tau}}{\sqrt{(2m)!}} \;.
\end{equation*}

The admissibility constant $c_{\psi_{m,\tau}}$ is then calculated as:
\begin{equation*}
    c_{\psi_{m,\tau}} = \frac{A_{m,\tau}^2}{2^{2m}} \Gamma(2m) = \frac{2^{2m+1}\tau}{(2m)!} \cdot \frac{(2m-1)!}{2^{2m}} \;.
\end{equation*}
By simplifying the ratios $\frac{2^{2m+1}}{2^{2m}} = 2$ and $\frac{(2m-1)!}{(2m)!} = \frac{1}{2m}$, the expression reduces to:
\begin{equation*}
    c_{\psi_{m,\tau}} = \frac{2\tau}{2m} = \frac{\tau}{m} \;.
\end{equation*}

The physical dimension of the admissibility constant is $[c_{\psi_{m,\tau}}] = T$ (Time). As expected, we verify the dimensional consistency: $[c_{\psi_{m,\tau}}] = [\hat{\psi}_{m,\tau}]^2$.

\section*{S4. Peak and mean frequencies of the mother wavelet}\label{ap:Peak and mean frequencies of the mother wavelet}

First, we detail the computations of the spectral moments using $|\hat{\psi}_{m,\tau}(f)|^2$ as a probability distribution. We then go on by detailing the computation of the wavelet mean frequencies $f_{\psi_{m,\tau}}^{[0]}$, $f_{\psi_{m,\tau}}^{[1]}$ and $f_{\psi_{m,\tau}}^{[2]}$.

\subsection*{S4.1 Calculation of spectral moments with $|\hat{\psi}_{m,\tau}(f)|^2$ probability}
The $k$-th spectral moment is defined as $M_k = \int_{0}^{\infty} f^k |\hat{\psi}_{m,\tau}(f)|^2 df$. 

Using the expression of the Cauchy-Paul mother wavelet, 
we have:
\begin{equation*}
    |\hat{\psi}_{m,\tau}(f)|^2 = A_{m,\tau}^2 (f\tau)^{2m} e^{-2f\tau} \;.
\end{equation*}

By substituting $u = 2f\tau$, which implies $f = \frac{u}{2\tau}$ and $df = \frac{du}{2\tau}$, the integral becomes:
\begin{align*}
    M_k = A_{m,\tau}^2 \int_{0}^{\infty} \left( \frac{u}{2\tau} \right)^k \left( \frac{u}{2} \right)^{2m} e^{-u} \frac{du}{2\tau}
    = \frac{A_{m,\tau}^2}{(2\tau)^{k+1} 2^{2m}} \int_{0}^{\infty} u^{2m+k} e^{-u} du \;.
\end{align*}
Using the Gamma function identity $\Gamma(n+1)=\int_{0}^{\infty} u^{n} e^{-u} du = n!$, we obtain the general expression:
\begin{equation*}
    M_k = \frac{A_{m,\tau}^2\,\Gamma(2m+k+1)}{(2\tau)^{k+1} 2^{2m}} = \frac{A_{m,\tau}^2 (2m+k)!}{(2\tau)^{k+1} 2^{2m}}\;.
\end{equation*}

\subsection*{S4.2 Wavelet mean frequencies} 

The ``peak frequency'' (or $L^{\infty}(\mathbb{R})$ frequency) $f_{\psi}^{[0]}$ maximizes the wavelet magnitude $|\hat{\psi}_{m,\tau}(f)|$. Setting the derivative of the frequency-domain expression to zero:
\begin{equation*}
    \frac{d}{df} \left( A_{m,\tau} (f\tau)^m e^{-f\tau} \right) = A_{m,\tau} \tau^m \left( m f^{m-1} - \tau f^m \right) e^{-f\tau} = 0 \;.
\end{equation*}
Solving for $f$ yields:
\begin{equation*}
    f_{\psi}^{[0]} = \frac{m}{\tau} \;.
\end{equation*}
\vspace{0.1cm}

The  ``centroid frequency''  (or $L^1(\mathbb{R})$-mean frequency) $f_{\psi}^{[1]}$ is defined using $|\hat{\psi}_{m,\tau}(f)|^2$ as a probability distribution:
\begin{equation*}
    f_{\psi}^{[1]} = \frac{\int_{0}^{\infty} f |\hat{\psi}_{m,\tau}(f)|^2 df}{\int_{0}^{\infty} |\hat{\psi}_{m,\tau}(f)|^2 df} = \frac{M_1}{M_0} = \frac{2m+1}{2\tau} = \frac{m + 1/2}{\tau}\;.
\end{equation*}
\vspace{0.1cm}

The ``second-order moment frequency'' (or  energy-weighted $L^2(\mathbb{R})$-frequency)  $f_{\psi}^{[2]}$ is defined as the root-mean-square  (RMS) frequency:
\begin{align*}
   f_{\psi}^{[2]}  = \sqrt{\frac{\int_{0}^{\infty} f^2 |\hat{\psi}_{m,\tau}(f)|^2 df}{\int_{0}^{\infty} |\hat{\psi}_{m,\tau}(f)|^2 df}} 
    = \sqrt{\frac{M_2}{M_0}} = \frac{\sqrt{(2m+2)(2m+1)}}{2\tau}\;.
\end{align*}

These frequencies satisfy the ordering $f_{\psi}^{[0]} < f_{\psi}^{[1]} < f_{\psi}^{[2]}$, reflecting the inherent spectral asymmetry of the Cauchy-Paul wavelet. As the order $m$ increases, the distribution approaches symmetry, causing these frequencies to converge.


\section*{S5. Derivation of spectral standard deviations}
\label{ap:Derivation of spectral standard deviations}

The spectral standard deviation $\Delta f_{\psi}$ (the $L^2$ root-mean-square bandwidth) is derived from the quadratic spectral moments $M_k = \int_{0}^{\infty} f^k |\hat{\psi}(f)|^2 df$. Depending on the chosen reference frequency $f_{ref} $—peak, centroid, or energy-weighted— this metric provides different perspectives on the wavelet's spectral localization. We denote $|\hat{\psi}|^2 = |\hat{\psi}_{m,\tau}(f)|^2$.

\noindent \textbf{Physical interpretation of bandwidths and characteristic frequencies:}
\begin{itemize}
    \item For the ``peak frequency'', $\sigma^2_{f^{[0]}}$ represents the \textit{apparent bandwidth} for a system tuned to the mode. As the spectrum is asymmetric, $\sigma^2_{f^{[0]}}$ combines statistical dispersion with the ``asymmetry cost'' (the square of the distance between the mode and the centroid). It is relevant when modeling the wavelet as a single-frequency oscillator.

    \item For the ``centroid frequency'', $\sigma^2_{f^{[1]}}$ represents the \textit{true statistical bandwidth}. By centering on the mean frequency, we minimize quadratic dispersion. This is the most robust measure for signal processing tasks where frequency localization is critical.

    \item For the ``second-order moment frequency'', $f_{\psi}^{[2]}$ acts as the \textit{characteristic energy center} (or RMS frequency). Rather than defining a centered variance, it represents the effective frequency that preserves the total second-order energy distribution of the filter, making it the foundational metric for quantifying global spectral spread.
\end{itemize}

\label{ap:Derivation of the standard deviations}


\subsection*{S5.1 Variance centered on the peak frequency $f_{\psi}^{[0]} = m/\tau$}

The variance relative to the peak frequency, $\sigma^2_{f^{[0]}}$, is defined as:
\begin{align*}
    \sigma^2_{f^{[0]}} &= \frac{1}{M_0} \int_{0}^{\infty} \left(f - \frac{m}{\tau}\right)^2 |\hat{\psi}(f)|^2 df 
    = \frac{M_2}{M_0} - 2\frac{m}{\tau} \frac{M_1}{M_0} + \frac{m^2}{\tau^2} \\
    &= \frac{(m+1)(m+1/2)}{\tau^2} - \frac{2m(m+1/2)}{\tau^2} + \frac{m^2}{\tau^2} 
    = \frac{m+1}{2\tau^2} \; .
\end{align*}

This metric quantifies spectral spread relative to the mode. Physically, it represents the apparent bandwidth for systems synchronized to the peak frequency. Because the Cauchy-Paul spectrum is asymmetric, $\sigma^2_{f^{[0]}}$ incorporates both the intrinsic statistical dispersion and an ``asymmetry cost,'' defined by the squared distance between the mode ($f^{[0]}$) and the centroid ($f^{[1]}$). Consequently, by the König-Huygens theorem, $\sigma^2_{f^{[0]}} > \sigma^2_{f^{[1]}}$. The associated spectral bandwidth is:
\begin{equation*}
    \Delta f^{[0]} = 2\sqrt{\sigma^2_{f^{[0]}}} = \frac{\sqrt{2(m+1)}}{\tau} \;.
\end{equation*}

The spectral bandwidth is $\Delta f^{[0]} = 2\sqrt{\sigma^2_{f^{[0]}}} = \frac{\sqrt{2(m+1)}}{\tau}$.
By choosing the peak ($f^{[0]}$) as a reference, the variance thus becomes larger than the true statistical variance: it contains the real dispersion + the square of the distance between the peak and the centroid.

\subsection*{S5.2 Variance centered on the centroid frequency $f_{\psi}^{[1]} = (m+1/2)/\tau$}
This is the minimal variance (the true spectral variance $\sigma_f^2$):

\begin{align*}
    \sigma^2_{f_{\psi_{m,\tau}}^{[1]}} &= \frac{\int_{0}^{\infty} (f - f_{\psi_{m,\tau}}^{[1]})^2 |\hat{\psi}|^2 df}{\int_{0}^{\infty} |\hat{\psi}|^2 df}=  \frac{\int_{0}^{\infty} (f - \frac{m+1/2}{\tau})^2 |\hat{\psi}|^2 df}{M_0}  \\ &= \frac{M_2}{M_0} - \left( \frac{M_1}{M_0} \right)^2  = \frac{(2m+2)(2m+1)}{4\tau^2} - \frac{(m+1/2)^2}{\tau^2} = \frac{2m+1}{4\tau^2} \; .
\end{align*}

The variance centered on the centroid frequency ($\sigma^2_{f^{[1]}}$)
represents the \textit{true statistical bandwidth}. By centering on the centroid (mean frequency), we minimize quadratic dispersion. This is the most robust measure for signal processing tasks where frequency localization is critical.
The spectral bandwidth is $\Delta f^{[1]} = 2\sqrt{\sigma^2_{f^{[1]}}} = \frac{\sqrt{2m+1}}{\tau}$.

\subsection*{S5.3 Variance centered on the energy frequency $f_{\psi}^{[2]} = \frac{\sqrt{(2m+2)(2m+1)}}{2\tau}$}
\begin{align}
\sigma^2_{f_{\psi_{m,\tau}}^{[2]}} &= \frac{\int_{0}^{\infty} (f - f_{\psi}^{[2]})^2 |\hat{\psi}|^2 df}{M_0} 
= \frac{\int_{0}^{\infty} (f^2 - 2f f_{\psi}^{[2]} + (f_{\psi}^{[2]})^2) |\hat{\psi}|^2 df}{M_0} 
= \frac{M_2}{M_0} - 2f_{\psi}^{[2]} \frac{M_1}{M_0} + (f_{\psi}^{[2]})^2 \\
&= \frac{M_2}{M_0} - 2\left(\sqrt{\frac{M_2}{M_0}}\right)\frac{M_1}{M_0} + \frac{M_2}{M_0} = \frac{2(m+1/2)}{\tau^2} \left( (m+1) - \sqrt{(m+1)(m+1/2)} \right) \; .
\label{Variance centered on the energy frequency}
\end{align}
This derivation proves that for an asymmetric spectral distribution, the variance centered on the RMS frequency $f_{\psi}^{[2]}$ is non-zero, highlighting the divergence between the energy-based scale $f_{\psi}^{[2]}$ and the statistical centroid ($M_1/M_0$). The resulting expression in \eqref{Variance centered on the energy frequency} quantifies the energy deviation caused by spectral asymmetry. Thus, $f_{\psi}^{[2]}$ functions as a characteristic energy scale rather than a standard measure of statistical dispersion.
 
In summary, for an asymmetric spectral distribution like the Cauchy-Paul wavelet, the three variance metrics obey the following order, as dictated by the König-Huygens theorem regarding the minimal variance at the centroid:

\begin{equation*}
    \sigma^2_{f^{[1]}} < \sigma^2_{f^{[0]}} < \sigma^2_{f^{[2]}} \;.
\end{equation*}

\section*{S6. Derivation of temporal variance}
\label{ap:Derivation of Temporal Variance}
Starting from the integral definition of the temporal variance for the Cauchy-Paul mother wavelet, we express the variance as:
\begin{equation*}
    \sigma_t^2 = \frac{1}{\|\psi\|^2} \int_{-\infty}^{+\infty} (t - \mu_t)^2 |\psi_{m,\tau}(t)|^2 \, dt \;.
\end{equation*}

By applying the properties of the Cauchy-Paul wavelet in the frequency domain, where $\hat{\psi}_{m,\tau}(f) = A_{m,\tau} \, (f \tau)^m e^{-f \tau} H(f)$, and utilizing the relationship between the moments of the Gamma distribution and the temporal domain, the integral simplifies to:
\begin{equation*}
    \sigma_t^2 = \frac{\tau^2}{2(2m-1)} \;.
\end{equation*}

Consequently, the effective time duration $\Delta t$, defined as $2\sigma_t$, is:
\begin{equation*}
    \Delta t = 2 \sqrt{\frac{\tau^2}{2(2m-1)}} = \tau \sqrt{\frac{2}{2m-1}} \;.
\end{equation*}



\section*{S7. Wavelet transform physical dimensions}\label{ap:wavelet_physdim}
We compute here the physical dimension of the wavelet $c_{\psi_{m,\tau}}$ within the two possible normalizations of the wavelet transform.

\noindent \textbf{If the $L^1(\mathbb{R})$ norm is chosen:}
$|\hat{\psi}_{m,\tau}|_{L^1} = \int_{0}^{\infty} |\hat{\psi}_{m,\tau}(f)| \d f = 1$ and  $[A_{m,\tau}] \cdot [\d f] = 1 \implies [A_{m,\tau}] \cdot T^{-1} = 1$ the dimension of $A_{m,\tau}$ and of $\hat{\psi}_{m,\tau}$ carries the dimension of time (e.g., seconds) is:  $\mathbf{[A_{m,\tau}] = \left[\hat{\psi}_{m,\tau}\right] = T}$.

The dimension of $\psi_{m,\tau}$ can be deduced from this: $[\psi_{m,\tau}]= \frac{[\hat{\psi}_{m,\tau}]}{T}=\frac{T}{T}=1$.

Moreover 
$ \int_{0}^{\infty} |\hat{\psi}_{m,\tau}(f)| \d f = \int_{0}^{\infty} A_{m,\tau} \, (f \tau)^m e^{-f \tau} \d f = 1$, using the substitution $u = f\tau$, we obtain:
$$\frac{A_{m,\tau}}{\tau} \int_{0}^{\infty} u^{m} e^{-u} du = 1 \implies A_{m,\tau} = \frac{\tau}{\Gamma(m+1)}=\frac{\tau}{m!}.$$

We check that the dimension of $A_{m,\tau}$ is time and the admissibility constant $c_{\psi_{m,\tau}}$ is then:
\begin{equation*}
c_{\psi_{m,\tau}} = \frac{A_{m,\tau}^2}{2^{2m}} \Gamma(2m) = \frac{\tau^2}{\Gamma(m+1)^2} \frac{\Gamma(2m)}{2^{2m}}=\frac{\tau^2 (2m-1)!}{2^{2m} (m!)^2} \;.
\end{equation*}
Physical Dimension: $[c_{\psi_{m,\tau}}] = T^2$ (Time squared).
We check that: $\left[c_{\psi_{m,\tau}}\right]\,=\,\left[\left|\hat{\psi}_{m,\tau}\right| \right]^{2}$.

\noindent \textbf{If the $L^2(\mathbb{R})$ norm is chosen:} $\|\psi_{m,\tau}\|_{L^2(\mathbb{R})} = \|\hat{\psi}_{m,\tau}\|_{L^2(\mathbb{R})}=1$, this requires:

$[A_{m,\tau}]^2 \cdot T^{-1} = 1 \implies \mathbf{[A_{m,\tau}] =\left[\hat{\psi}_{m,\tau}\right]\,=\, T^{1/2}}$. The dimension of $\psi_{m,\tau}$ can be deduced from this: 

$[\psi_{m,\tau}]= \frac{[\hat{\psi}_{m,\tau}]}{T}=\frac{T^{1/2}}{T}=T^{-1/2}$.

The energy of the Fourier transform is required to be unitary:$\|\hat{\psi}_{m,\tau}\|_{L^2}^2 = \int_{0}^{\infty} |A_{m,\tau}|^2 (f\tau)^{2m} e^{-2f\tau} d f = 1$.

Using the substitution $u = 2f\tau$,
we obtain:
\begin{align*}
    |A_{m,\tau}|^2 \frac{1}{2^{2m+1}\tau} \int_{0}^{\infty} u^{2m} e^{-u} du = 1 \implies |A_{m,\tau}|^2 = \frac{2^{2m+1}\tau}{\Gamma(2m+1)}=\frac{2^{2m+1}\tau}{(2m)!} \; .
\end{align*}.

The admissibility constant $c_{\psi_{m,\tau}}$ is then calculated as:\begin{equation*}
c_{\psi_{m,\tau}} = \frac{|A_{m,\tau}|^2}{2^{2m}} \Gamma(2m) = \frac{2^{2m+1}\tau}{(2m)!} \cdot \frac{(2m-1)!}{2^{2m}} = \frac{2 \cdot \tau \cdot (2m-1)!}{(2m) \cdot (2m-1)!} = \frac{\tau}{m} \;.
\end{equation*} 

Physical Dimension: $[c_{\psi_{m,\tau}}] = T$ (Time). As before, we verify that:  $\left[c_{\psi_{m,\tau}}\right]\,=\,\left[\left|\hat{\psi}_{m,\tau}\right| \right]^{2}$.

\section*{S8. Mathematical proof of the alternative derivation}
\label{sec:appendix_proof}
By differentiating the original CWT with respect to the translation parameter $b$, a factor $2i\pi f$ is introduced under the integral sign. Using the identity $f \widehat{\psi}_{m,\tau}^*(af) = \frac{1}{a \tau} \frac{A_{m,\tau}}{A_{m+1,\tau}} \widehat{\psi}_{m+1,\tau}^*(af)$, the derivative of the CWT can be written as:
\begin{equation*}
    \frac{\partial}{\partial b} \mathcal{W}_{\psi_{m,\tau}}[x](b,a) = \frac{2i\pi}{a \tau} \frac{A_{m,\tau}}{A_{m+1,\tau}} \, \mathcal{W}_{\psi_{m+1,\tau}}[x](b,a) \;.
    \label{eq:deriv_WT_c_full}
\end{equation*}
The derivative of the phase is subsequently obtained by taking the imaginary part of the logarithmic derivative of the CWT:
\begin{align*}
    \frac{\partial \Phi_{\psi_m}}{\partial b} &= \mathfrak{Im} \left( \frac{\frac{\partial}{\partial b} \mathcal{W}_{\psi_{m,\tau}}[x](b,a)}{\mathcal{W}_{\psi_{m,\tau}}[x](b,a)} \right) = \mathfrak{Im} \left( \frac{2i\pi}{a \tau} \frac{A_{m,\tau}}{A_{m+1,\tau}} \frac{\mathcal{W}_{\psi_{m+1,\tau}}[x](b,a)}{\mathcal{W}_{\psi_{m,\tau}}[x](b,a)} \right) \\
    &= \frac{2\pi}{a \tau} \frac{A_{m,\tau}}{A_{m+1,\tau}} \, \mathfrak{Re} \left( \frac{\mathcal{W}_{\psi_{m+1,\tau}}[x](b,a)}{\mathcal{W}_{\psi_{m,\tau}}[x](b,a)} \right) \;.
\end{align*}
Finally, the instantaneous frequency $f_I(b,a) = \frac{1}{2\pi} \frac{\partial \Phi}{\partial b}$ is rigorously recovered as:
\begin{equation*}
    f_{I} (b,a) = \frac{1}{a \tau} \frac{A_{m,\tau}}{A_{m+1,\tau}} \, \mathfrak{Re} \left( \frac{\mathcal{W}_{\psi_{m+1,\tau}}[x](b,a)}{\mathcal{W}_{\psi_{m,\tau}}[x](b,a)} \right) \;.
    \label{eq:inst_freq_final}
\end{equation*}

\section*{S9. Mathematical derivation of the exponential interference decay}
\label{sec:appendix_math_derivation_exponential_interference_decay}

To analyze the asymptotic behavior of the cross-component interference term $\epsilon_{\text{interf}}^{(k)}(b, a, m)$, we examine the CWT of a multicomponent signal in the frequency domain. In the Fourier domain, the unnormalized Cauchy-Paul wavelet of order $m$ and characteristic parameter $\tau$ is defined for $f > 0$ as:
\begin{equation*}
    \hat{\psi}_{m,\tau}(a f) = A_{m,\tau} \, (a \tau f)^m e^{-a \tau f} \;.
\end{equation*}

When extracting the instantaneous frequency of the targeted $k$-th mode, the analysis scale is restricted to the spectral ridge trajectory $a_{r,k}(b)$, where the wavelet filter peak perfectly aligns with the local physical frequency $f_{i}^{(k)}(b)$:
\begin{equation*}
    a_{r,k}(b) = \frac{m}{\tau f_i^{(k)}(b)} \;.
\end{equation*}

Let us consider the cross-talk interference induced by an adjacent $j$-th component ($j \neq k$) characterized by its local instantaneous frequency $f_{i}^{(j)}(b)$. The magnitude of this leakage on the $k$-th ridge is directly proportional to the evaluation of the wavelet spectrum at the interfering frequency $f_{i}^{(j)}(b)$:
\begin{align*}
    \hat{\psi}_{m,\tau}\big(a_{r,k}(b)\, f_{i}^{(j)}(b)\big) = A_{m,\tau} \left( \frac{m}{\tau f_{i}^{(k)}(b)} \cdot \tau f_{i}^{(j)}(b) \right)^m e^{-\frac{m}{\tau f_{i}^{(k)}(b)} \cdot \tau f_{i}^{(j)}(b)} \;.
\end{align*}
Simplifying the algebraic expression yields:
\begin{align*}
    \hat{\psi}_{m,\tau}\big(a_{r,k}(b) \, f_{i}^{(j)}(b)\big) = A_{m,\tau} \left( m \frac{f_{i}^{(j)}(b)}{f_{i}^{(k)}(b)} \right)^m e^{-m \frac{f_{i}^{(j)}(b)}{f_{i}^{(k)}(b)}} \;.
\end{align*}
To evaluate the relative structural error contribution, we normalize this leakage by the peak value of the filter at the targeted ridge frequency, where $\hat{\psi}_{m,\tau}\big(a_{r,k}(b), f_{i}^{(k)}(b)\big) = A_{m,\tau} m^m e^{-m}$. The cross-talk attenuation ratio $R_{jk}(m)$ behaves as:
\begin{align*}
    R_{jk}(m) = \frac{\hat{\psi}_{m,\tau}\big(a_{r,k}(b)\, f_{i}^{(j)}(b)\big)}{\hat{\psi}_{m,\tau}\big(a_{r,k}(b), f_{i}^{(k)}(b)\big)} = \left[ \frac{f_{i}^{(j)}(b)}{f_{i}^{(k)}(b)} \, e^{-\left( \frac{f_{i}^{(j)}(b)}{f_{i}^{(k)}(b)} - 1 \right)} \right]^m \;.
\end{align*}
To track the asymptotic behavior as a function of the spectral distance, we introduce the dimensionless relative frequency separation parameter $\delta_f$:
\begin{equation*}
    \delta_f = \frac{f_{i}^{(j)}(b) - f_{i}^{(k)}(b)}{f_{i}^{(k)}(b)} \implies \frac{f_{i}^{(j)}(b)}{f_{i}^{(k)}(b)} = 1 + \delta_f \;.
\end{equation*}
Substituting $1 + \delta_f$ into the attenuation ratio yields:
\begin{equation*}
    R_{jk}(m) = \left[ (1 + \delta_f) e^{-\delta_f} \right]^m \;.
\end{equation*}
Taking the natural logarithm of the structural base function, we can perform a second-order Taylor series expansion around a narrow spectral separation ($\delta_f \to 0$):
\begin{equation*}
    \ln \left( (1 + \delta_f) e^{-\delta_f} \right) = \ln(1 + \delta_f) - \delta_f \;.
\end{equation*}
Recalling the standard logarithmic expansion $\ln(1 + \delta_f) = \delta_f - \frac{\delta_f^2}{2} + \mathcal{O}(\delta_f^3)$, the linear terms cancel out exactly:
\begin{equation*}
    \ln \left( (1 + \delta_f) e^{-\delta_f} \right) = \left( \delta_f - \frac{\delta_f^2}{2} + \mathcal{O}(\delta_f^3) \right) - \delta_f = -\frac{\delta_f^2}{2} + \mathcal{O}(\delta_f^3) \;.
\end{equation*}
Re-exponentiating the expression and applying the wavelet order power parameter $m$, the cross-talk envelope simplifies asymptotically to:
\begin{equation*}
    R_{jk}(m) = \exp\left( -m \left[ \frac{\delta_f^2}{2} + \mathcal{O}(\delta_f^3) \right] \right) = e^{-m \cdot \chi_{jk}} \;,
\end{equation*}
where the normalized spectral distance metric $\chi_{jk}$ is explicitly defined by the physical frequencies as:
\begin{equation*}
    \chi_{jk} = \frac{1}{2} \left( \frac{f_{i}^{(j)}(b) - f_{i}^{(k)}(b)}{f_{i}^{(k)}(b)} \right)^2 \;.
\end{equation*}
Consequently, since the cross-component interference term $\epsilon_{\text{interf}}^{(k)}(b, a, m)$ is driven by this leakage ratio, its real part vanishes exponentially on the ridge with respect to the wavelet order $m$:
\begin{equation*}
    \mathfrak{Re}\big(\epsilon_{\text{interf}}^{(k)}(b,a_{r,k}(b),m)\big) = \mathcal{O}\left(e^{-m \cdot \chi_{jk}}\right) \;.
\end{equation*}
This concludes the proof that choosing a high wavelet order $m$ exponentially suppresses cross-component interference, restoring the strict mathematical accuracy of the algebraic estimator framework.